%% file: ms.tex
\documentclass[twocolumn]{aastex63}

\shorttitle{Planetary Nebulae toward the Galactic Center}
\shortauthors{Hong et~al.}

\begin{document}

\title{Hunting for Planetary Nebulae toward the Galactic Center}

\author{Jihye Hong}
\affiliation{Department of Science Education, Ewha Womans University, 52 Ewhayeodae-gil, Seodaemun-gu, Seoul 03760, Republic of Korea; deokkeun@ewha.ac.kr}

\author{Janet P.\ Simpson}
\affiliation{SETI Institute, 189 Bernardo Avenue, Mountain View, CA 94043, USA}

\author{Deokkeun An}
\affiliation{Department of Science Education, Ewha Womans University, 52 Ewhayeodae-gil, Seodaemun-gu, Seoul 03760, Korea; deokkeun@ewha.ac.kr}

\author{Angela S.\ Cotera}
\affiliation{SETI Institute, 189 Bernardo Avenue, Mountain View, CA 94043, USA}

\author{Solange V. Ram{\'{\i}}rez}
\affiliation{Carnegie Observatory, 813 Santa Barbara Street, Pasadena, CA 91101, USA }


\begin{abstract}

We present near-infrared (IR) spectra of two planetary nebula (PN) candidates in close lines of sight toward the Galactic center (GC) using the Gemini Near-Infrared Spectrograph (GNIRS) at Gemini North. High-resolution images from radio continuum and narrow-band IR observations reveal ringlike or barrel-shaped morphologies of these objects, and their mid-IR spectra from the {\it Spitzer} Space Telescope exhibit rich emission lines from highly-excited species such as [\ion{S}{4}], [\ion{Ne}{3}], [\ion{Ne}{5}], and [\ion{O}{4}]. We also derive elemental abundances using the Cloudy synthetic models, and find an excess amount of the $s$-process element Krypton in both targets, which supports their nature as PN. We estimate foreground extinction toward each object using near-IR hydrogen recombination lines, and find significant visual extinctions ($A_V > 20$). The distances inferred from the size versus surface brightness relation of other PNe are $9.0\pm1.6$~kpc and $7.6\pm1.6$~kpc for SSTGC~580183 and SSTGC~588220, respectively. These observed properties along with abundance patterns and their close proximity to Sgr~A$^*$ (projected distances $\la20$~pc) make it highly probable that these objects are the first confirmed PN objects in the nuclear stellar disk. The apparent scarcity of such objects resembles the extremely low rate of PN formation in old stellar systems, but is in line with the current rate of the sustained star formation activity in the Central Molecular Zone.

\end{abstract}


\section{Introduction}

Most stars with initial masses less than $\sim 8\ M_\odot$ evolve into planetary nebulae (PNe) at the end of their lifetimes \citep[see][and references therein]{balick:02}, as long as their progenitor masses are large enough \citep[see][]{jacoby:97}. In our Galaxy, a large number of PNe ($\sim 3,500$) have been discovered so far, many of which are associated with the Galactic bulge or thin disk \citep[e.g.,][]{parker:06,miszalski:08,jacoby:10,sabin:14}, serving as bright tracers for kinematics \citep[e.g.,][]{durand:98,beaulieu:00} and chemical abundance studies in the Milky Way \citep[e.g.,][]{stanghellini:18}. However, it has been known for a while that this number is an order of magnitude lower than expected from population synthesis models \citep[e.g.,][]{moe:06}, and the majority of PN populations in the Milky Way remain to be discovered. The heavy dust obscuration near the Galactic plane is likely the main cause of such discrepancy \citep[e.g.,][]{jacoby:04,miszalski:08,parker:12}.

In this context, the absence of PNe in the nuclear bulge of the Milky Way \citep{serabyn:96,launhardt:02} can be understood to be the result of the large amount of foreground dust toward the Galactic center (GC; $A_V\sim30$), which essentially prohibits detections and identifications of PNe through the conventional method of measuring optical emission lines. The mass of the nuclear bulge is $\sim1.4\times10^9\ M_\odot$ \citep{launhardt:02}, which is approximately $10$ times less massive than the kiloparsec-scale classical bulge. The stellar populations in the nuclear bulge are predominantly old \citep[e.g.,][]{nogueras:20}, but may have distinct chemical properties from the classical bulge \citep[e.g.,][]{schultheis:20}. Most of these stars are confined to the nuclear stellar disk, a rotating disk of stars around Sgr~A$^*$. The nuclear stellar disk spatially overlaps with the Central Molecular Zone \citep[CMZ; see][and references therein]{morris:96}, a massive reservoir of molecular gas clouds with a diameter of $\sim500$~pc and a total cloud mass of $\sim3$--$8\times10^7\ M_\odot$ \citep{dahmen:98,tsuboi:99}. Sustained star formation activity is observed throughout the region \citep[e.g.,][]{an:11,longmore:13}. Given that both the Galactic bulge and thin disk harbor a noticeable number of PNe, the lack of PNe in the nuclear bulge poses a challenge to our understanding of PN formation, stellar populations, and evolution in the innermost region of the Milky Way.

Recently, we have identified two objects, SSTGC~580183 (G359.9627-0.1202) and SSTGC~588220 (G0.0967-0.0511), as candidate PNe \citep{simpson:18} while analyzing mid-infrared (IR) spectra of compact IR sources toward the CMZ \citep{an:09,an:11}. They show high-excitation lines such as [\ion{Na}{3}] $7.3\ \mu$m, [\ion{Ne}{5}] $14.3\ \mu$m and $24.3\ \mu$m, and [\ion{O}{4}] $25.9\ \mu$m. These lines are not observed in typical \ion{H}{2} regions, but are often seen in PNe with an extremely hot ($3\times10^4 \la T_{\rm eff} \la 2\times10^5$~K) central object \citep[e.g.,][]{osterbrock:06,peimbert:17}. The two objects have also been observed in high-resolution radio observations and with near-IR narrowband filters, revealing a ringlike or barrel-shaped nebulosity \citep{wang:10,zhao:20}.

In this paper, we report near-IR spectroscopic follow-up observations of these two objects using the Gemini Near-Infrared Spectrograph \citep[GNIRS;][]{elias:06a,elias:06b} at the Frederick C. Gillett Gemini North Telescope. Owing to a paucity of PNe toward the GC, both objects are unique, providing an opportunity to explore stellar evolution and chemical characteristics of stellar populations in the inner region of the Galaxy. Their probable location in the CMZ is an interesting aspect of this study, since the massive reservoir of molecular gas in the CMZ is known to maintain the most intense star-forming activity in the Milky Way \citep[$\sim0.1\ M_\odot\ yr^{-1}$;][]{an:11,longmore:13}; they can be used to study chemical and kinematical properties of stars that are distinct from those of other Galactic components.

The primary goal of this study is to confirm the nature of these nebular sources as PNe and to characterize their chemical properties based on near-IR and mid-IR spectra. The near-IR spectra are particularly useful for this purpose, because hydrogen recombination lines can be used to constrain foreground extinction, which is necessary to derive the elemental abundances of ionic species from mid-IR spectra. This paper is organized as follows. In \S~\ref{sec:observation}, we describe the observing tactics and data reduction. In \S~\ref{sec:results}, we present the GNIRS spectra, derive foreground extinction from hydrogen recombination lines, and put a constraint on their distances from a comparison with size versus surface brightness relations of PNe. We conduct a joint analysis of near-IR and mid-IR spectra in \S~\ref{sec:cloudy} to derive elemental abundances. A summary of the work is provided in \S~\ref{sec:summary}.

\begin{figure*}[!th]
\epsscale{2.1}
\plottwo{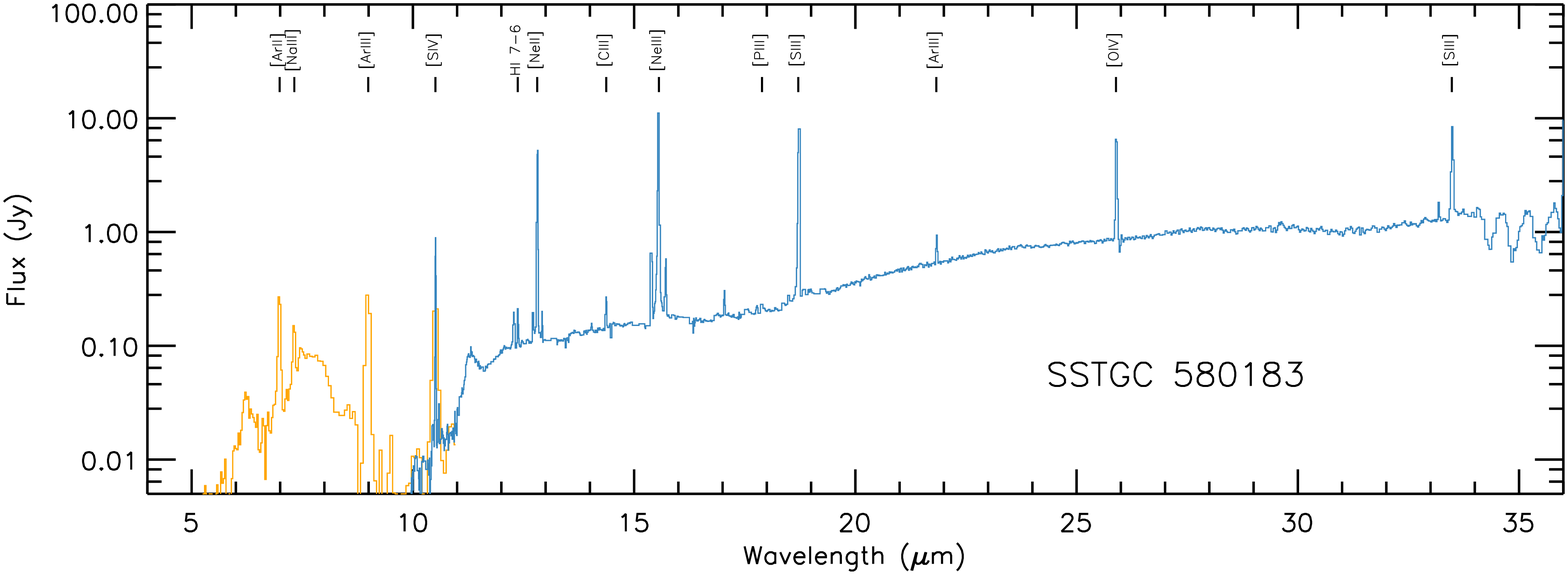}{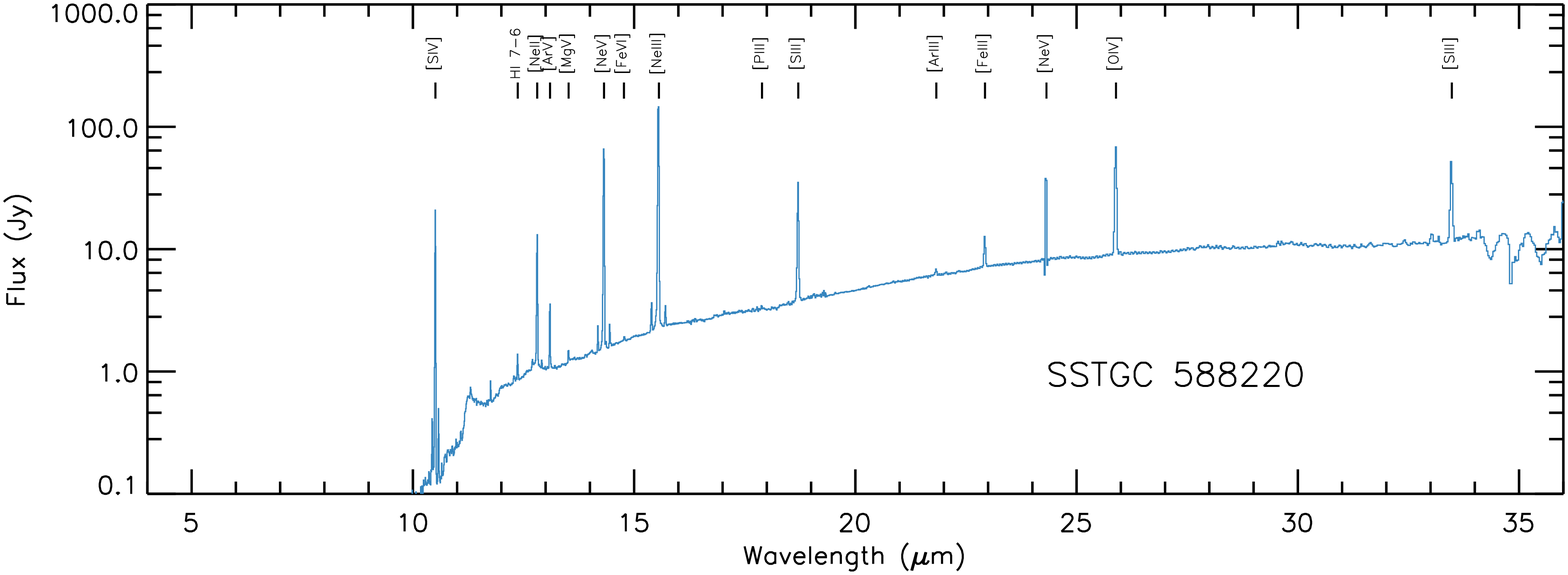}
\caption{Background-subtracted {\it Spitzer}/IRS spectra of PN candidates. Spectra from high-resolution modules are shown by blue lines. For SSTGC~580183, low-resolution spectra are displayed by orange lines.}
\label{fig:irs}
\end{figure*}

\begin{figure*}[!th]
\epsscale{1}
\plottwo{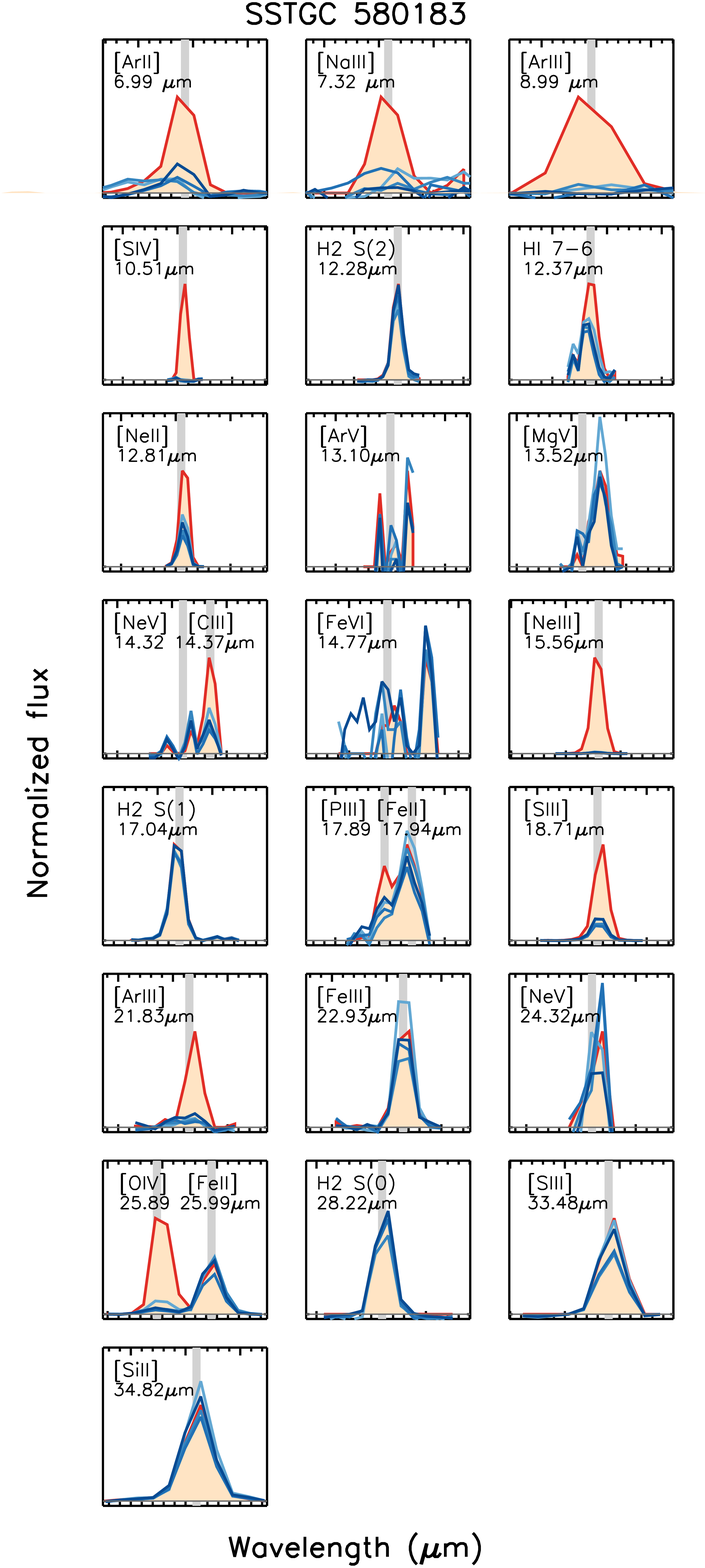}{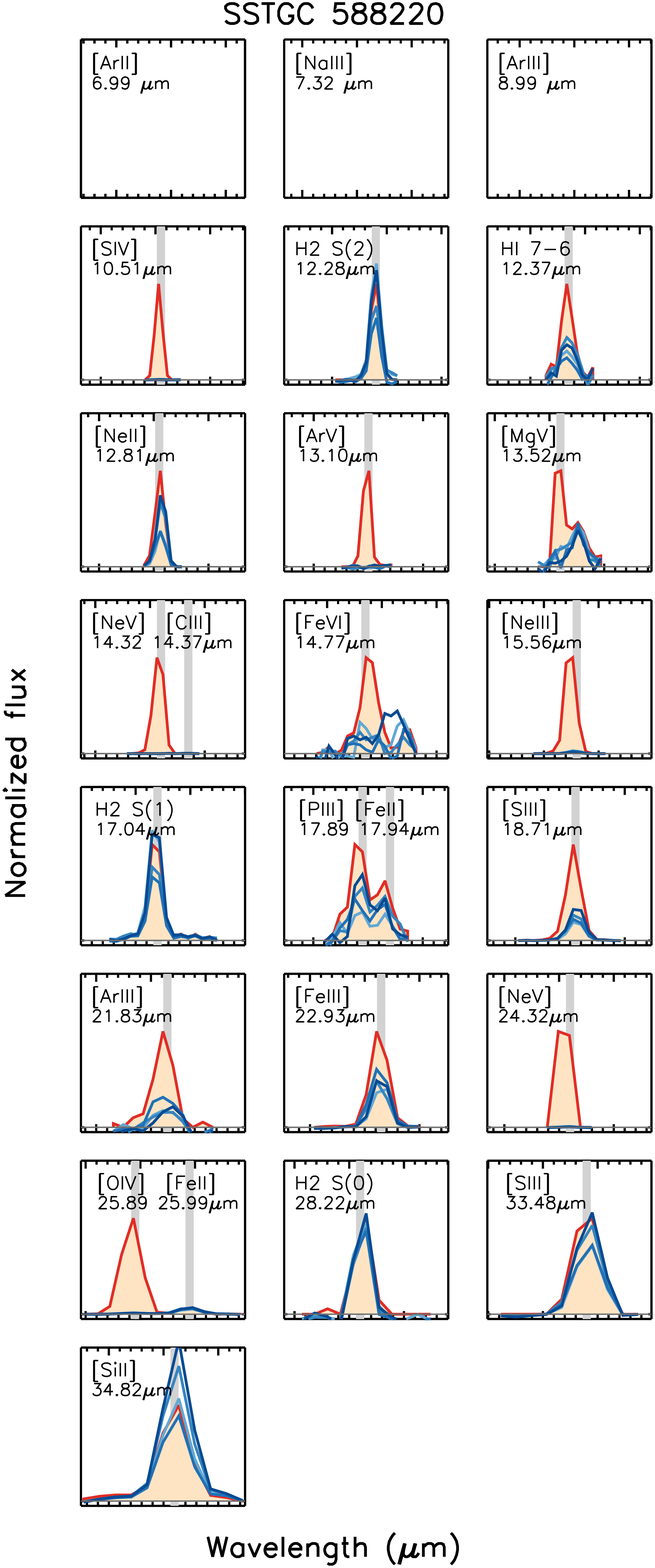}
\caption{Emission lines in the {\it Spitzer}/IRS spectra of the PN candidates. Original line emissions from each target are shown by the orange-shaded red lines, while the blue lines show emission lines extracted at four nearby locations that do not contain bright point sources in the slit. Line centers are marked by a vertical gray stripe. Each panel spans a wavelength range of $0.3\ \mu$m (each tick mark covers $\Delta\lambda=0.02\ \mu$m).} 
\label{fig:irs2}
\end{figure*}

\section{Observations and Data Reduction}\label{sec:observation}

\subsection{Candidates}

Both SSTGC~580183 ($\alpha=17$h $46$m $0.034$s, $\delta=-29\arcdeg$ $1\arcmin$ $50.210\arcsec$) and SSTGC~588220 ($\alpha$=$17$h $46$m $2.984$s, $\delta=-28\arcdeg$ $52\arcmin$ $45.320\arcsec$) were originally identified as compact (within a $2\arcsec$ beam) sources in {\it Spitzer} Space Telescope \citep{werner:04} Infrared Array Camera \citep[IRAC;][]{fazio:04} images \citep{ramirez:08}. They were targeted for follow-up observation using the Infrared Spectrograph \citep[IRS;][]{houck:04} on board {\it Spitzer} \citep{an:09,an:11}, as part of a search for massive young stellar objects in the CMZ. There are no parallax measurements, but their proximity to the GC (with a projected angular distance of $\sim7\arcmin$) is deemed as a supporting piece of evidence for their potential membership in the CMZ, which covers a region of $\sim3\arcdeg\times0.5\arcdeg$ centered at the GC.

Figure~\ref{fig:irs} shows mid-IR spectra of each target, taken using the high-resolution modules of IRS \citep[see][for more details]{an:09,an:11}. The background emission from surrounding clouds in the CMZ was subtracted from the spectrum of each target using a set of background spectra (see below). The continuum of the background spectra is characterized by warm-dust emission and strong emission from polycyclic aromatic hydrocarbons at $6.2$, $7.7$, $11.3$, $12.7$, and $16.4\ \mu$m; nonetheless, they are $\sim2$--$3$ times fainter than the target spectrum in the short-high (SH) module ($10\ \mu$m $\la \lambda \la 20\ \mu$m), and $\sim5$--$8$ times fainter in the long-high (LH) module ($20\ \mu$m $\la \lambda \la 30\ \mu$m). There are also a number of strong emission lines seen in the background spectra, which originate from diffuse ionized gas and photodissociation regions (PDRs) in molecular clouds adjacent to each target.

The mean background spectrum of the high-resolution modules was constructed using IRS observations at four carefully chosen locations ($\sim1\arcmin$ away in each direction) with the same instrument setup. Figure~\ref{fig:irs2} shows the emission lines observed in the IRS spectra after the background subtraction \citep[see][for more information on the IRS observations]{an:11}. The strong PDR emission lines from H$_2$ S(2) $12.28$ and H$_2$ S(1) $17.04\ \mu$m are negligible after the background subtraction. On the other hand, while some forbidden emission lines from highly-excited ion species such as [\ion{S}{4}] $10.51$ and [\ion{O}{4}] $25.89\ \mu$m are not visible or are very weak in the background spectra, they are clearly seen in the background-subtracted spectra.\footnote{[\ion{O}{4}] $25.89\ \mu$m is saturated in the high-resolution IRS spectrum of SSTGC~588220, for which we used the low-resolution IRS spectrum in the modeling (see below).} SSTGC~588220 additionally exhibits emission from [\ion{Ar}{5}] 13.10~\micron, [\ion{Mg}{5}] 13.52~\micron, and [\ion{Fe}{6}] 14.77~\micron. Such lines are commonly observed in PNe from gas ionized by a source of temperature $\sim10^5$~K, rather than from typical \ion{H}{2} regions ionized by massive OB stars \citep[e.g.,][]{osterbrock:06}.

\subsection{Gemini/GNIRS Observations}

\begin{figure}[!th]
\epsscale{2.3}
\plottwo{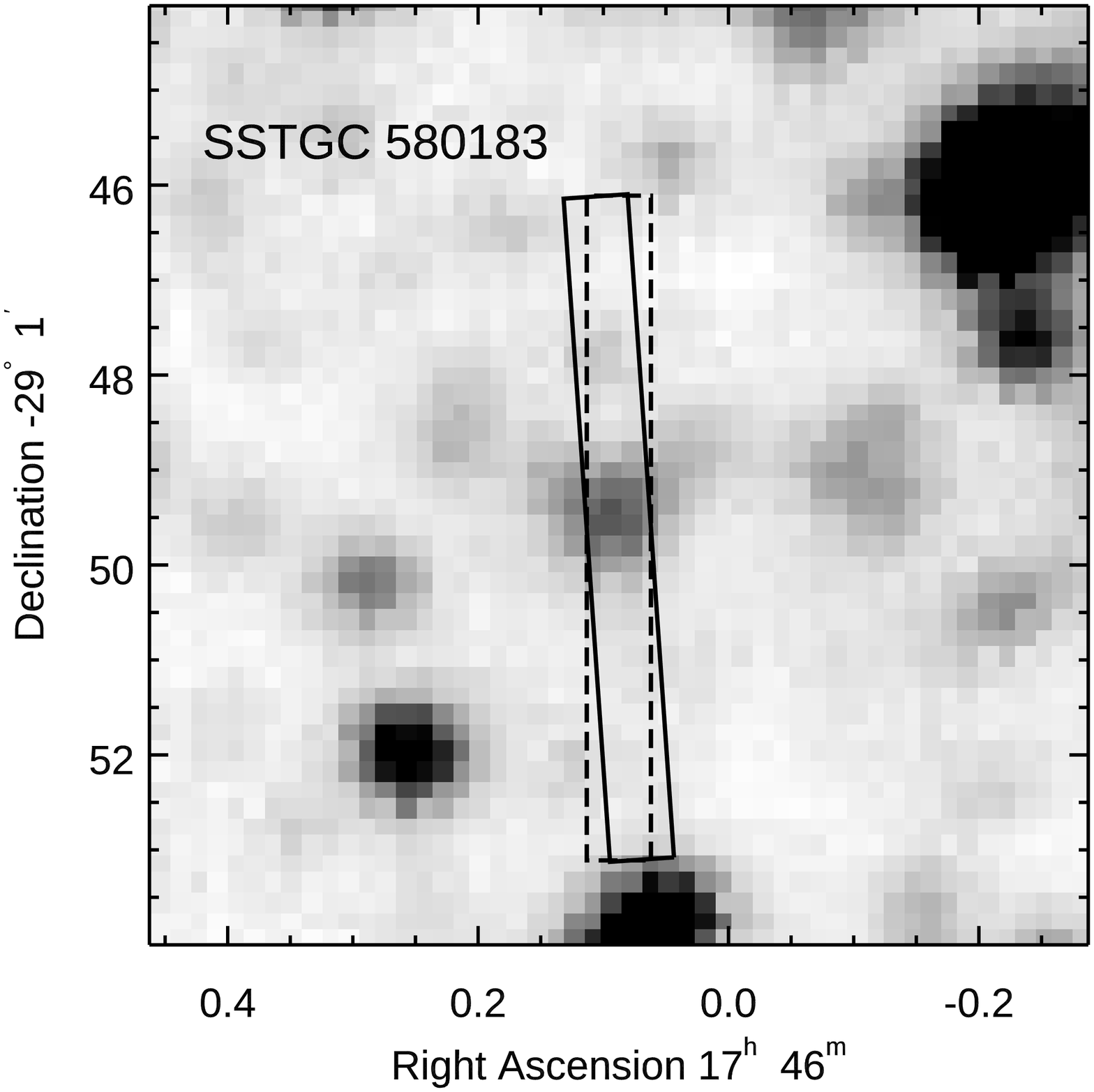}{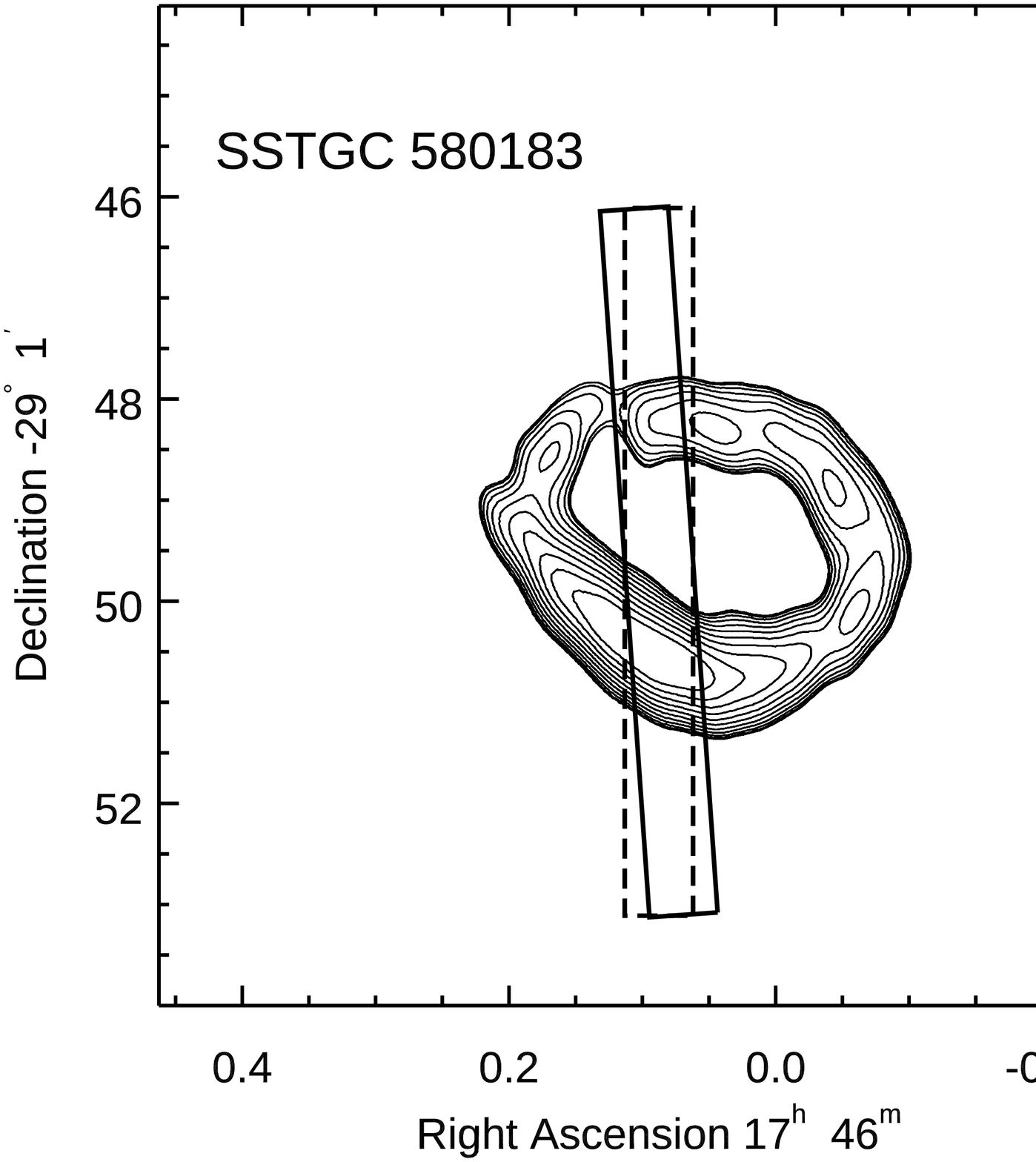}
\caption{The UKIDSS $K$-band image \citep[][top]{lawrence:07} and the JVLA $5.5$~GHz contour plot \citep[][bottom]{zhao:20} of SSTGC~580183. The GNIRS slit is overlaid on top of each image. The dashed and solid lines indicate the slit position/angle on 2016 March 11 and 2016 April 16, respectively. North is to the top, and east to the left.}
\label{fig:image}
\end{figure}

\begin{figure}[!th]
\epsscale{1.1}
\plotone{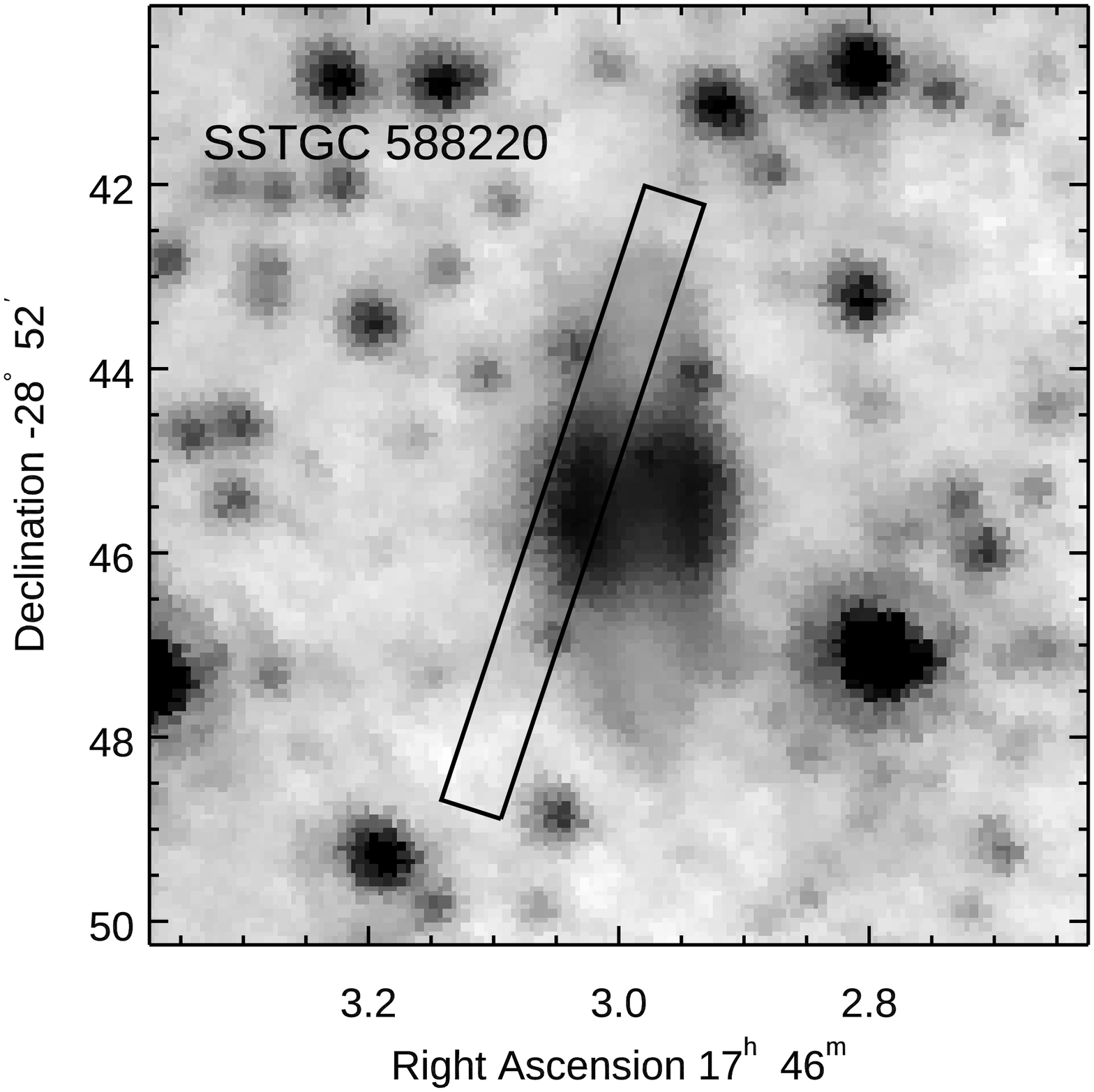}
\caption{{\it HST}/NICMOS Paschen $\alpha$ (F187) image \citep{wang:10} of SSTGC~588220. North is to the top, and east to the left. The long rectangle indicates the location of the slit on 2016 March 19.}
\label{fig:image2}
\end{figure}

Medium-resolution ($R\sim2600$) near-IR ($1.05\ \mu$m to $2.22\ \mu$m) spectra of the two PN candidates were taken using Gemini/GNIRS. A $0.675\arcsec \times 7\arcsec$ slit with the $110.5$ mm$^{-1}$ grating was used in cross-dispersed (XD) mode with the short-blue camera ($0.15$\arcsec per pixel); the slit width was set to approximately match the seeing during the observation. Both targets were observed in queue observing runs; SSTGC~580183 was observed twice, on 2016 March 11 and April 16, with the former observation taken with a nonzero parallactic angle. In the following analysis, we used averaged line fluxes from both observations, but took only the March data at short wavelengths ($\lambda < 2\ \micron$), because no such emission lines were detected in the latter data set. SSTGC~588220 was observed on 2016 March 19. The sky was mostly clear and the seeing was stable ($0.5\arcsec$--$0.7\arcsec$) during the observations.

Figure~\ref{fig:image} displays locations of GNIRS slits for SSTGC~580183, overlaid on top of the UKIRT Infrared Deep Sky Survey \citep[UKIDSS;][]{lawrence:07} $K$-band image (top panel) and a Jansky Very Large Array (JVLA) $5.5$~GHz radio image \citep[][bottom]{zhao:20}. This object was not resolved in earlier radio observations \citep{yusefzadeh:04,mills:11}, but the higher-resolution image in \citet{zhao:20} clearly reveals a ringlike structure with a diameter of $\sim4\arcsec$ along the major axis. The GNIRS slit was put nearly at the center of the ring.

Previously, SSTGC~588220 was observed in a Paschen-$\alpha$ imaging survey using the {\it Hubble Space Telescope} \citep{wang:10}. As displayed in Figure~\ref{fig:image2},  the high-resolution ($\sim0.2\arcsec$) image reveals a bright central region with a diameter of $\sim2\arcsec$ and a fainter elongated ring that extends $\sim3\arcsec$ from the center along the north-south direction. According to the morphological classification of PNe in \citet{sahai:11}, the object can be classified as having a pair of collimated lobes (L class; the fainter, elongated ringlike structure) with a barrel-shaped central region (brr(o) subclass). Numerical simulations in \citet{akashi:18} suggest that such barrel-like PNe can be formed through interactions of circumstellar material with jets in a binary system. As shown by the long rectangular box in Figure~\ref{fig:image2}, we put the GNIRS slit in a way that the center of the slit pass through the eastern part of the barrel-shaped central region. The position angle at the time of observation was set to the parallactic angle; fortuitously, a part of the northern (fainter) lobe was also observed and included in the following analysis.

Because of the high source density in the GC, the off-source slit position was carefully chosen near each target, and a background spectrum was obtained in an on-off sequence. The total on-source exposure time was $300$~s each night. We began reducing the GNIRS data by removing pattern noise and performing flat-field correction using the Gemini {\tt IRAF} packages\footnote{IRAF is distributed by the National Optical Astronomy Observatory, which is operated by the Association of Universities for Research in Astronomy (AURA) under a cooperative agreement with the National Science Foundation.}. For each set of data frames, we adjusted bias levels by matching the median values of pixels between spectral orders. We then followed the standard data reduction procedure for GNIRS\footnote{https://www.gemini.edu/instrumentation/gnirs/data-reduction} to correct for order distortion, and perform wavelength calibration using an argon lamp.

\begin{figure}
\epsscale{1.1}
\plotone{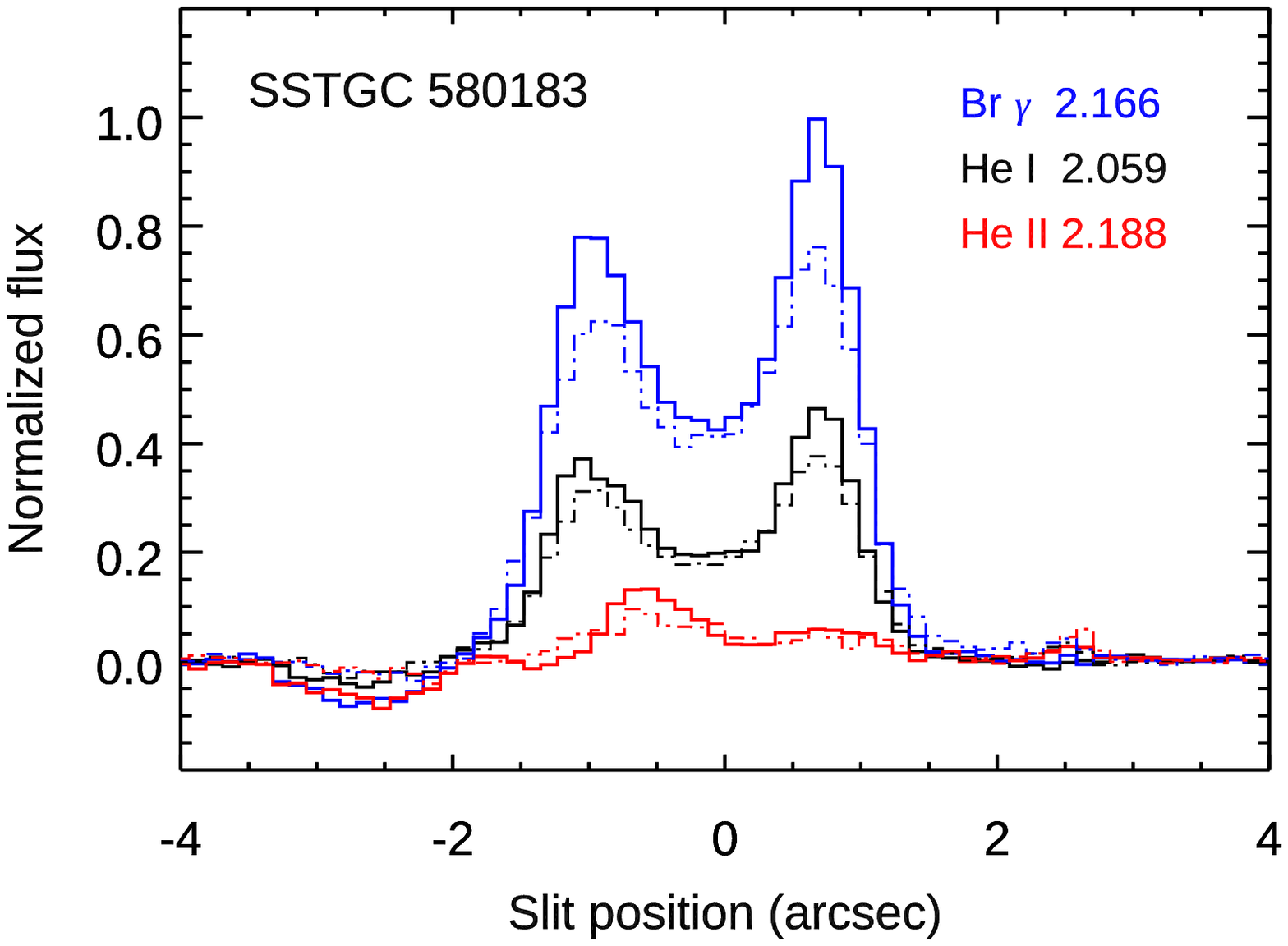}
\epsscale{1.2}
\plotone{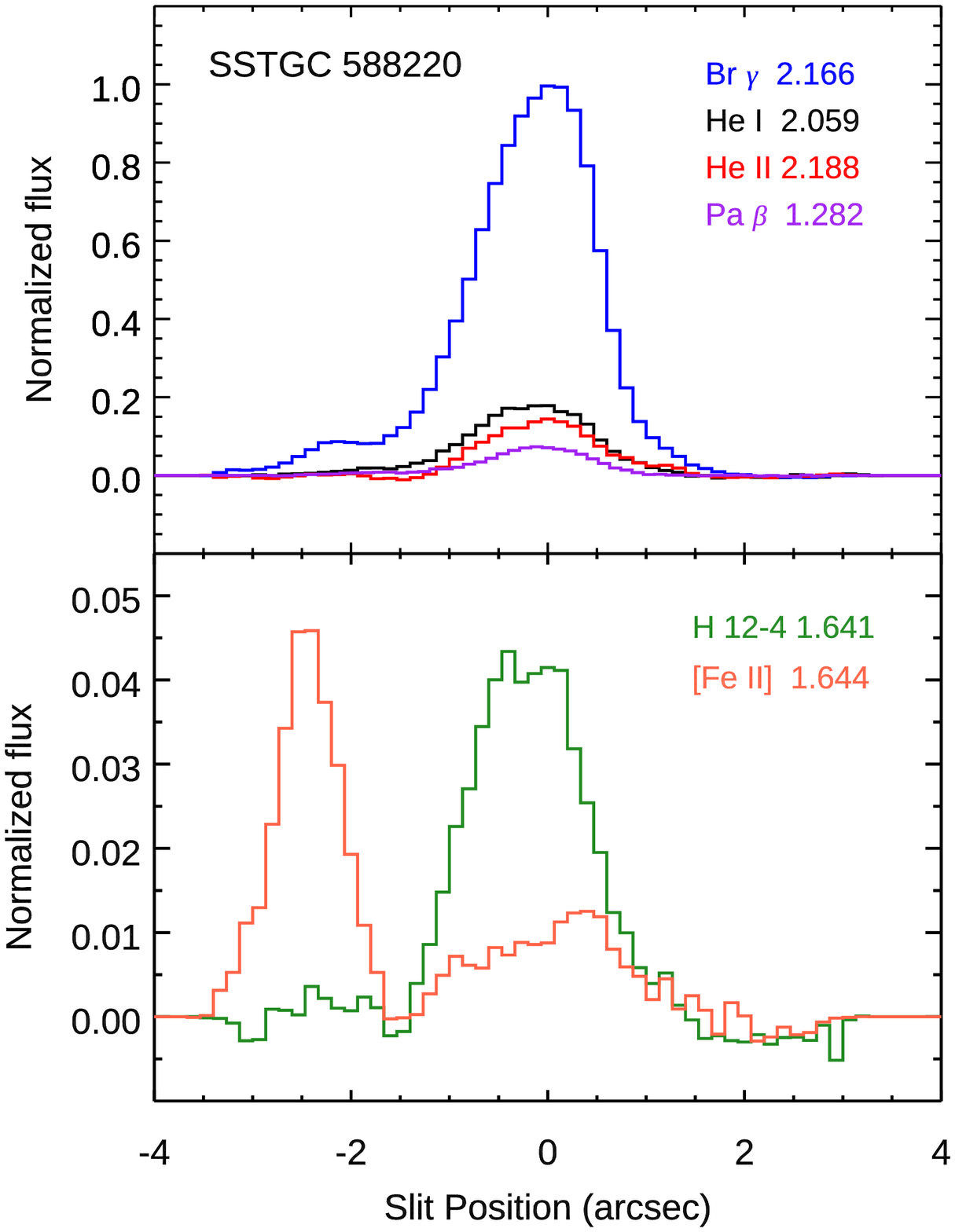}
\caption{The flux distribution of individual lines along the slit for SSTGC~580183 (top) and SSTGC~588220 (middle and bottom). Line fluxes summed over $6$~\AA\ are normalized with respect to the maximum value of Br~$\gamma$ emission. The dotted-dashed and solid lines in the top panel show the line fluxes from the first and second runs on the target, respectively. North is to the left.}
\label{fig:profile}
\end{figure}

Figure~\ref{fig:profile} shows source profiles along the slit. For SSTGC~580183 (top panel), a striking bimodal structure is seen in the line emission profiles; we attribute this to the ringlike morphology \citep[R class in][]{sahai:11} observed in the radio image (see the bottom panel of Figure~\ref{fig:image}). The bimodal structures in strong lines, such as Br~$\gamma$ and \ion{H}{1} 7--4 ($2.166\ \mu$m), are similar, and are also seen in weaker lines, although to a lesser degree. To collect light from the entire line emission structure, we used a $4.2\arcsec$-wide aperture for spectral extraction. 

On the other hand, SSTGC~588220 (middle and bottom panels of Figure~\ref{fig:profile}) shows a single peak, although the slit contains extended emission and its dispersion is larger than the seeing (FWHM$\sim0.6\arcsec$--$0.7\arcsec$). Most notably, [\ion{Fe}{2}] 1.644~\micron\ from SSTGC~588220 is significantly brighter in the northern lobe, and is displaced by $\sim2.5\arcsec$ from the central emission (bottom panel). Other lines also show emission at the location of the northern lobe, but its strength is significantly weaker than the one observed in the slit center. We used a $3.6\arcsec$ aperture to extract the spectra, collecting both the central emission and the emission from the northern lobe. For [\ion{Fe}{2}], the emission line spectrum was extracted at the place where the line emission was strongest. We corrected for telluric absorptions using Spextool v4.1 \citep{cushing:04}, based on nightly observations of the telluric standard star (HD~157918).

\section{Results}\label{sec:results}

\subsection{Line Flux Measurements}

\input{tab1.tex}

\begin{figure*}[!th]
\center
\includegraphics[scale=0.42]{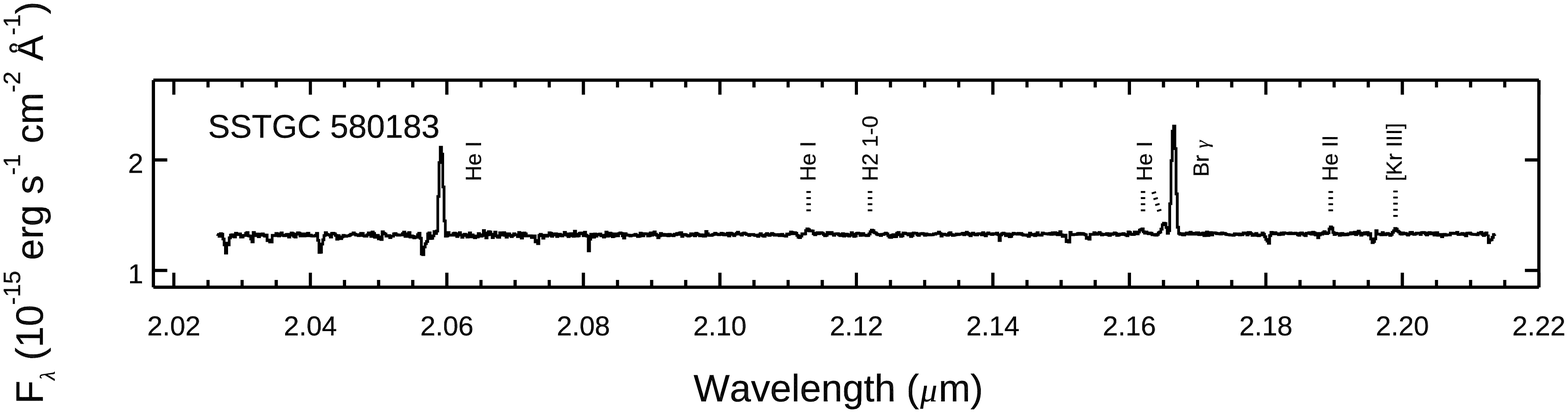}
\includegraphics[scale=0.36]{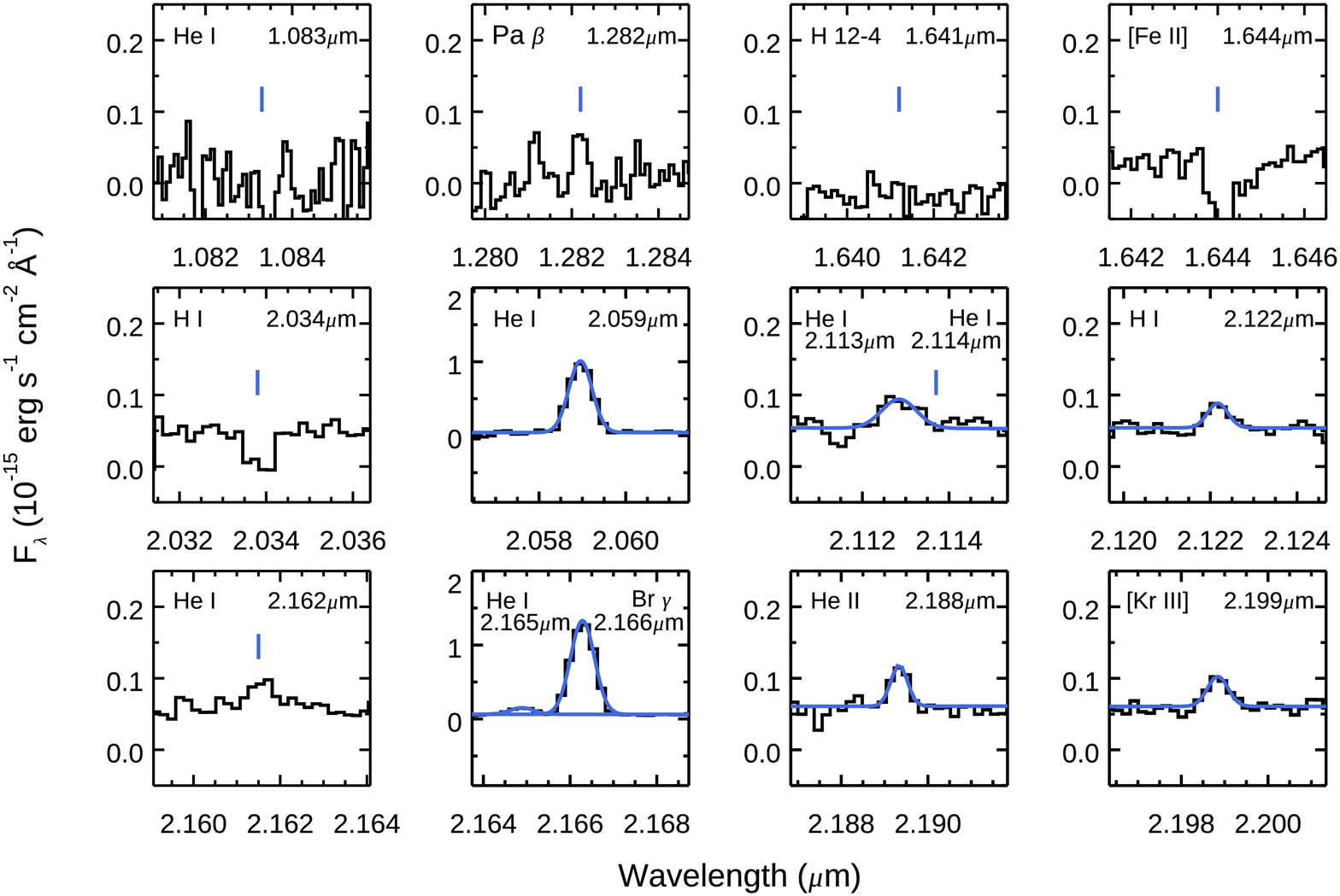}
\caption{GNIRS spectra of SSTGC~580183. The other spectral orders are omitted, because none of the emission lines are detected at shorter wavelengths.}
\label{fig:gnirs}
\end{figure*}

\begin{figure*}[!th]
\center
\includegraphics[scale=0.42]{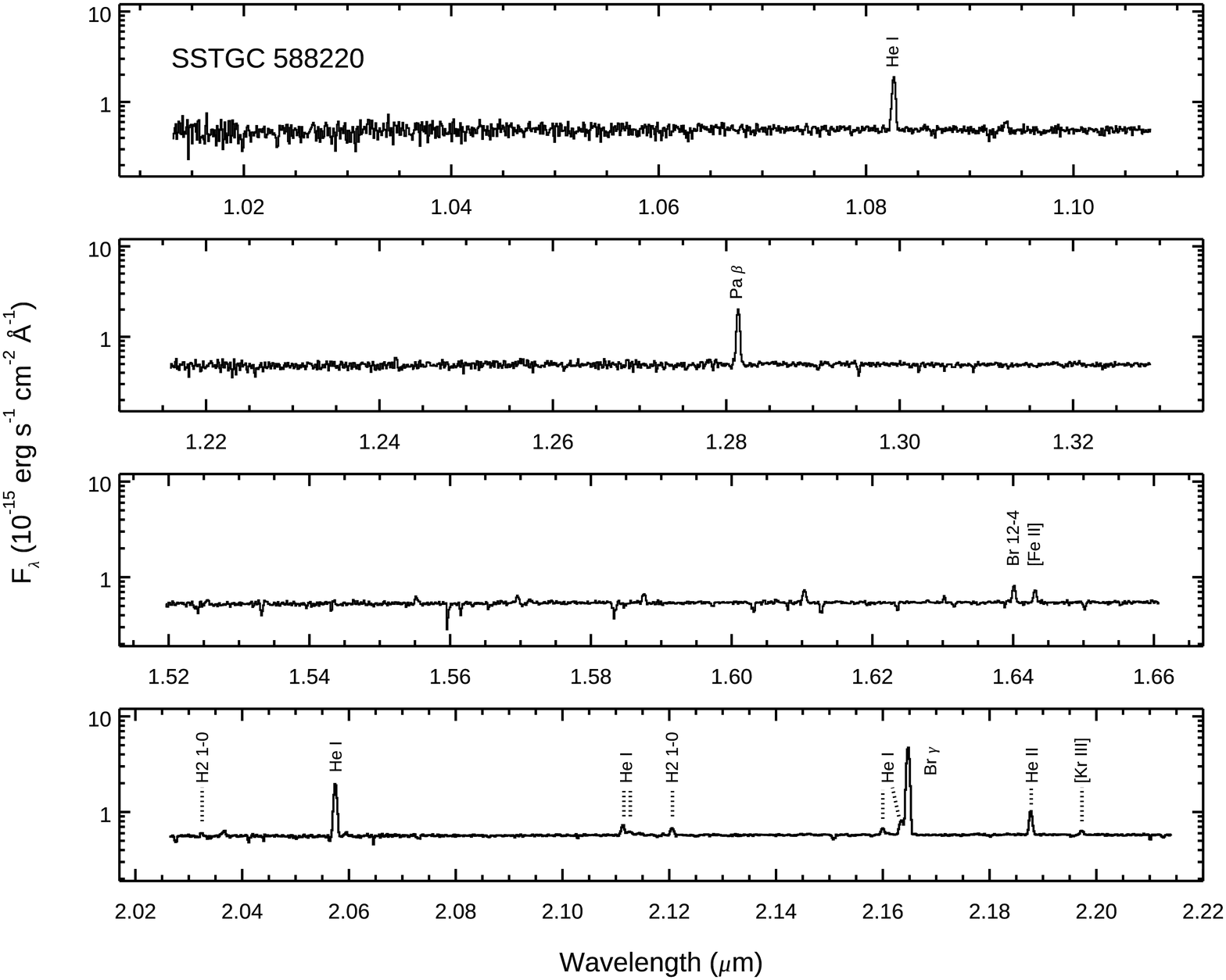}
\includegraphics[scale=0.36]{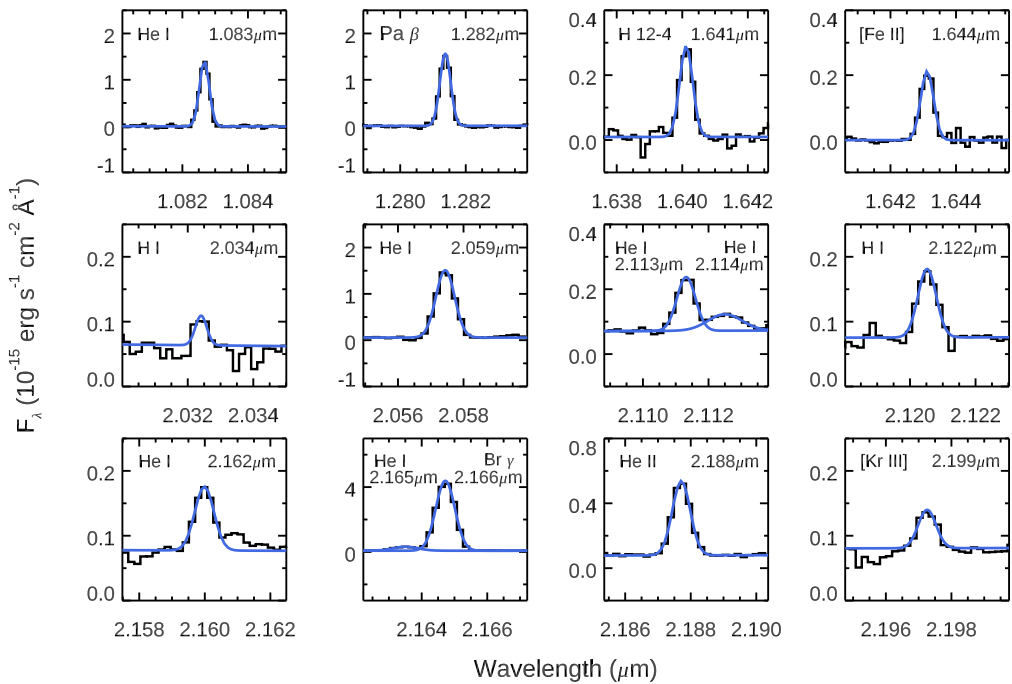}
\caption{GNIRS spectra of SSTGC~588220.}
\label{fig:gnirs2}
\end{figure*}

Our GNIRS spectra of SSTGC~580183 and SSTGC~588220 are shown in Figures~\ref{fig:gnirs} and \ref{fig:gnirs2}, respectively. The observed spectra contain several hydrogen and helium recombination lines, as well as [\ion{Fe}{2}] 1.644~\micron, [\ion{Kr}{3}] 2.199~\micron, H$_2$ 1--0 S(1) 2.1218~\micron, and S(2) 2.0338~\micron. Among these, Br$~\gamma$ ($2.166\ \mu$m) and Pa$~\beta$ ($1.282\ \mu$m) are of particular importance in this study, since they provide a strong constraint on the amount of foreground extinction toward each object. For SSTGC~580183, only the $K$-band spectrum from the longest wavelength order is shown, because most lines are essentially undetected (see individual panels at the bottom). Even [\ion{Fe}{2}] shows a possible absorption feature, which might reflect the widespread [\ion{Fe}{2}] emission from the interstellar medium (ISM) around the source \citep[see][]{an:13,simpson:18}. Pa~$\beta$ was detected at $2.5\sigma$ in March, but it was not seen in the April data due to a slightly lower signal-to-noise ratio (S/N). The weak Pa~$\beta$ is indicative of a large foreground extinction (see below). 

Table~\ref{tab:tab1} contains the line fluxes measured from the GNIRS spectra. They were measured by fitting a Gaussian function after subtracting the local continuum using a straight line. For blended lines such as \ion{He}{1} ($2.165\ \mu$m) and Br~$\gamma$, two Gaussians were used to simultaneously fit both emission lines. Flux uncertainties indicate the difference between the best-fitting Gaussian and a direct summation of the continuum-subtracted line fluxes. A $3\sigma$ upper limit is shown if the lines are not detected, as is the case for most of the short-wavelength lines in SSTGC~580183. The heliocentric radial velocities ($v_r^{\rm helio}$) are the flux-weighted, averaged values from the observed lines, and the uncertainties represent the standard deviation of these measurements. Radial velocities with respect to the local standard of rest (LSR) are also given ($v_r^{\rm LSR}$).

\subsection{Foreground Extinction Estimates}

As seen from Br~$\gamma$ emission, which is stronger than Pa~$\beta$, both objects suffer strong attenuation by a large amount of foreground dust in the Galactic disk. Such a high extinction implies a large distance from the Sun, as we discuss below. However, it also means that modeling emission lines is sensitively affected by the adopted foreground extinction. The impacts of extinction corrections on IR lines are not as significant as those required at optical wavelengths, but systematic differences in the IR extinction curves, as demonstrated below, and patchy extinction toward the CMZ generally make extinction corrections difficult.

\input{tab2.tex}

Table~\ref{tab:tab2} lists individual extinction estimates from the observed Br~$\gamma$ and Pa~$\beta$ lines for each object, based on three near-IR extinction curves \citep{chiar:06,boogert:11,fritz:11}. For the estimate derived from \citet{chiar:06}, we employed their extinction curve in the line of sight to the GC. The $3\sigma$ upper limit on Pa~$\beta$ was used to constrain the foreground extinction toward SSTGC~580183. We assumed the Case~B emissivity ratios from \citet{storey:95}, $j({\rm Pa}~\beta)/j({\rm Br}~\gamma)=5.875$, at a typical nebular electron temperature ($T_e=10^4$~K) and density ($N_e=10^3~{\rm cm}^{-3}$) \citep[e.g.,][]{zhang:04} and compared them to the observed line ratios to derive the extinction at $2.2\ \mu$m, $A_K$. The uncertainties are the quadratic sum of the propagated values from the flux measurement uncertainties and systematic differences from other Case~B conditions ($2\times10^3 \leq T_e \leq 2\times10^4$~K and $10^2 \leq N_e \leq 10^4~{\rm cm}^{-3}$).

As shown in Table~\ref{tab:tab2}, all of the three extinction curves produce large $A_K$ for both objects, as expected from the observed Br~$\gamma$ and Pa~$\beta$ line ratios. However, the exact values of the foreground extinction strongly depend on the adopted extinction curve. In particular, the curve of \citet{fritz:11} has the largest slope in the near-IR among the three curves, which results in a systematically smaller $A_K$ by $\sim30\%$. The differences among the three extinction laws exceed the observational uncertainties.

Nevertheless, the above extinction estimates for SSTGC~588220 are smaller than the {\it mean} extinction toward sources in the CMZ. The average foreground extinction toward the CMZ is $\langle A_V \rangle \approx 30$~mag, or $\langle A_K \rangle \approx 3.3$~mag, if $A_K/A_V=0.11$ is adopted \citep{figer:99}. The $A_K$ from \citet{fritz:11} is almost a factor of 2 lower than this average. In the case of SSTGC~580183, where we used a $3\sigma$ upper limit on Pa~$\beta$, the foreground extinction estimates are slightly larger than the GC average, except from the \citeauthor{fritz:11} curve.

In Table~\ref{tab:tab2}, {\it local} values of the mean foreground extinction are also included. \citet{schultheis:09} provided an extinction map across the CMZ, based on the IR colors of red giant branch stars in a grid of $2\arcmin\times2\arcmin$. The $A_K$ represents the mean and standard deviation of their measurements within $2\arcmin$ from each source. Taking $A_K/A_V=0.11$, their estimates correspond to $A_V\sim30$--$40$, which is consistent with the average extinction toward the GC. While the \citet{schultheis:09} map represented the bulk average of foreground extinction toward each source, \citet{an:13} provided local extinction that is more specific to their lines of sight. In practice, \citet{an:13} followed the procedure in \citet{simpson:07}, and estimated the optical depth at $9.6\ \mu$m ($\tau_{9.6}$) from the flux ratio between $10$ and $14\ \mu$m, measuring the strength of the silicate absorption band centered at $9.6\ \mu$m \citep[see also][]{simpson:18}. The $\langle \tau_{9.6} \rangle$ and its uncertainty are the average and standard deviation from four nearby {\it Spitzer}/IRS slits (within $0.9\arcmin$--$1.5\arcmin$ of the sources) that were originally designed to measure background CMZ emissions \citep{an:09,an:11}.

In contrast, \citet{simpson:18} estimated values of $\tau_{9.6} = 2.725$ for SSTGC~580183 and $2.803$ for SSTGC~588220, using a combination of the shapes of the $10\ \micron$ silicate feature and the [\ion{S}{3}] $18.7\ \micron/33.5\ \micron$ line ratios. We emphasize that these extinction values, like those from \citet{an:13}, apply to the diffuse ISM in the GC and not to objects much smaller than the {\it Spitzer} IRS apertures that may be substantially in front of (or behind) the CMZ of the GC.

The conversion between $A_K$ and $\tau_{9.6}$ also depends on the shape of the extinction curve over the wavelength interval. If $A_V/\tau_{9.6}=9$ \citep{roche:85} is taken, along with $A_K/A_V=0.11$, $\tau_{9.6}/A_K$ becomes unity. The two extinction curves employed in this work predict $\tau_{9.6}/A_K=1.3$ \citep{chiar:06} and $1.8$ \citep{fritz:11}, implying relatively strong silicate absorption with respect to $A_K$. In Table~\ref{tab:tab2}, $\langle \tau_{9.6} \rangle$ is the mean from the two cases with the \citet{chiar:06} and \citet{fritz:11} laws, while the uncertainty indicates half of the difference. The $\tau_{9.6}$ estimates from the GNIRS spectra are $15\%$--$20\%$ smaller than the ISM extinction measurements from \citet{an:13}, implying that the source is likely located in front of gas and dust in the CMZ. Taking these at face value, the larger $A_K$ from background giants \citep{schultheis:09} also implies a shorter distance to SSTGC~588220 than to the GC. In the case of SSTGC~580183, the $3\sigma$ upper limits are comparable to the foreground extinction from \citet{schultheis:09}, while they are higher than the \citet{an:13} estimates.

\subsection{Comparison to the Size vs. Surface Brightness Relations of PNe}\label{sec:size}

\begin{figure}
\centering
\includegraphics[scale=0.19]{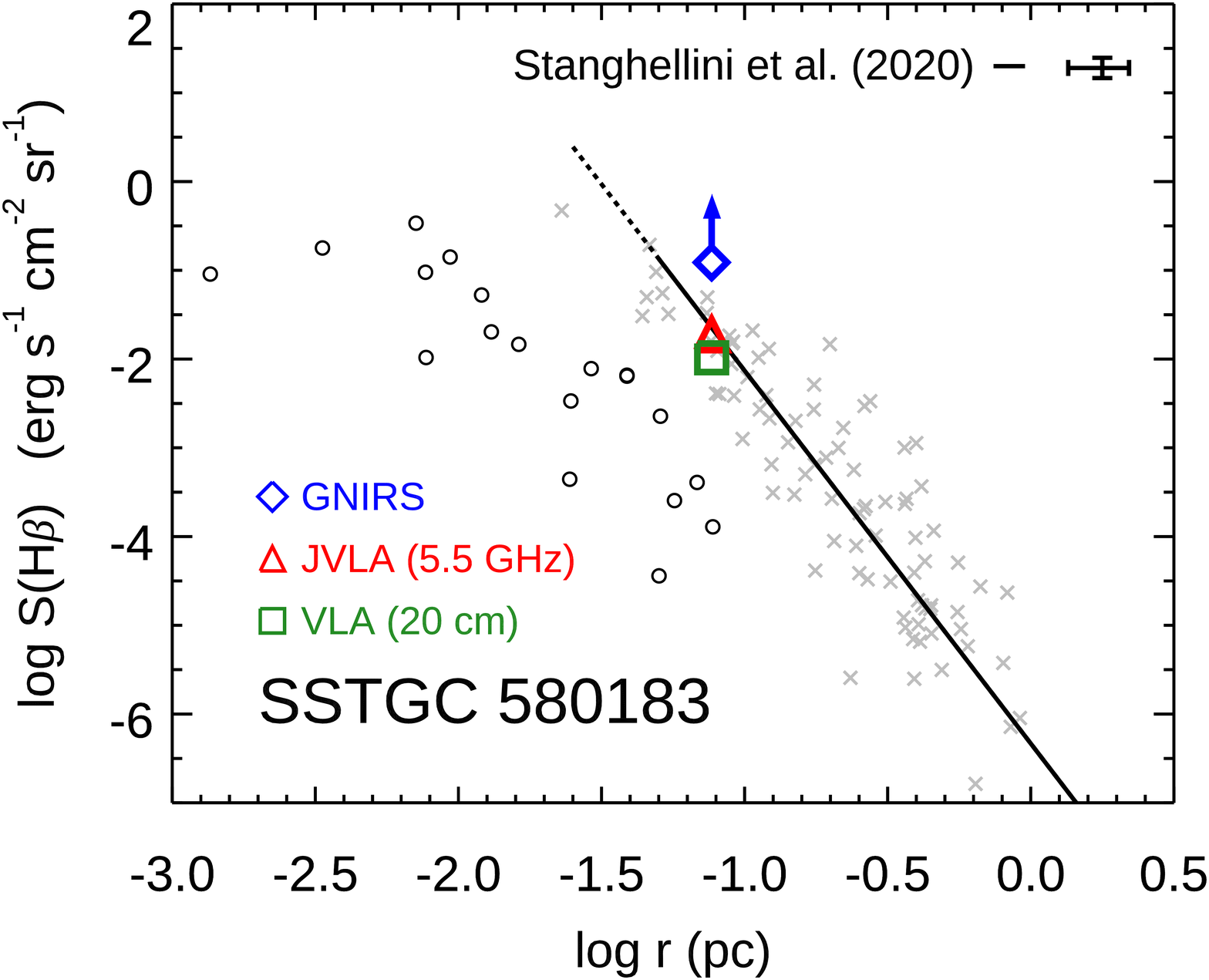}
\includegraphics[scale=0.19]{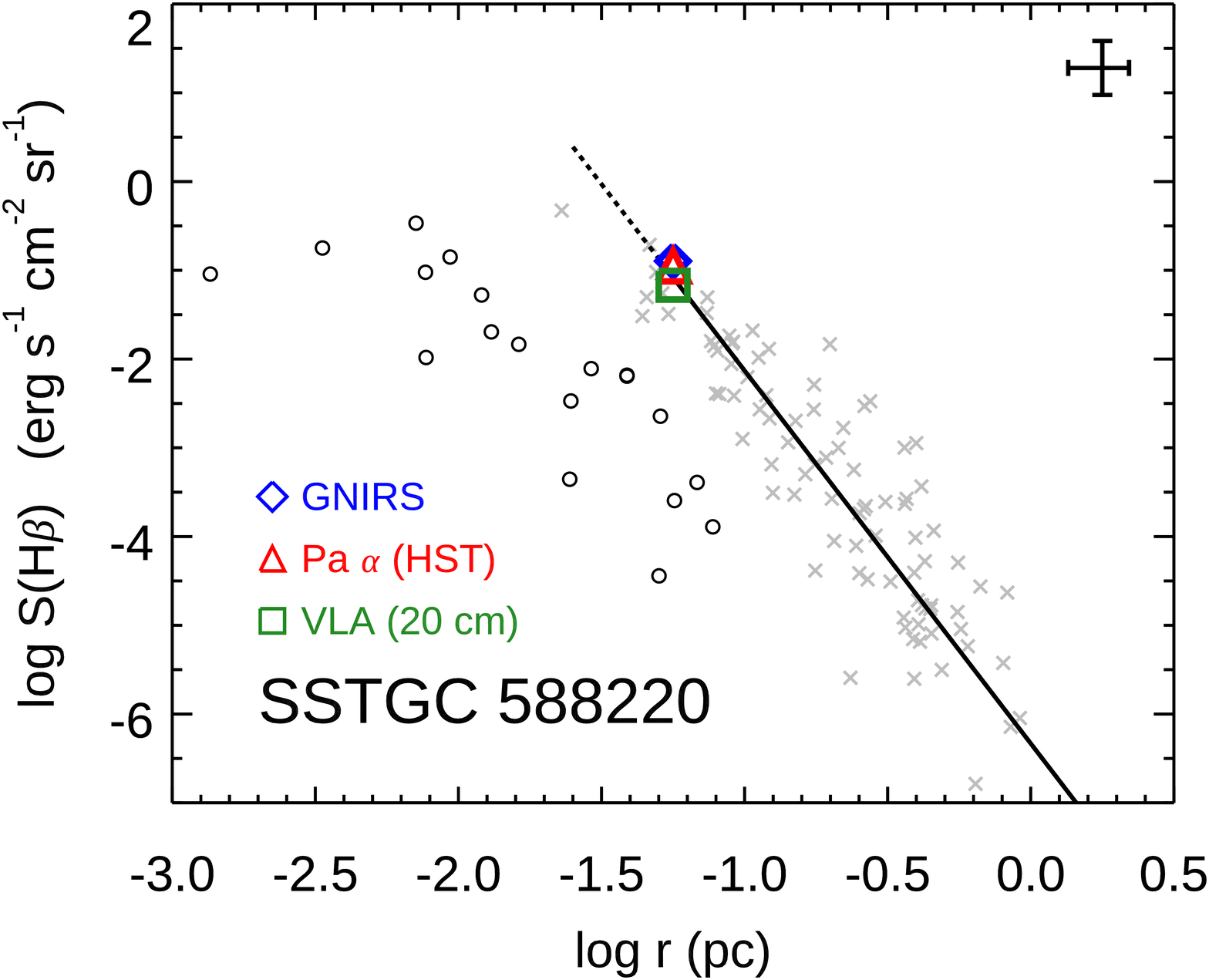}
\caption{Comparisons of SSTGC~580183 (top) and SSTGC~588220 (bottom) to the radius versus H$\beta$ surface brightness relation of PNe in \citet{stanghellini:20}. The surface brightness estimates from GNIRS observations are indicated by the blue open diamonds. Other measurements from JVLA 5.5~GHz \citep[][red triangle at the top]{zhao:20}, VLA 20~cm \citep[][green boxes]{yusefzadeh:04}, and {\it HST} Pa~$\alpha$ observations \citep[][red triangle at the bottom]{wang:10} are also shown assuming that the sources are located at the GC distance ($d_\odot=8.2$~kpc). For the purpose of comparison, representative error bars are shown for $\Delta d_\odot = \pm2$~kpc (see text). The GNIRS-based estimate for SSTGC~580183 indicates a $3\sigma$ lower limit. The gray crosses are Galactic PNe in \citet{stanghellini:20}, and the open circles indicate a subset of these with low ionized-mass (see their Figure~3); their best-fitting power law is shown by a solid line, and its extrapolation beyond the sample limit is indicated by a dotted line.}
\label{fig:stanghellini}
\end{figure}

\begin{figure}
\centering
\includegraphics[scale=0.19]{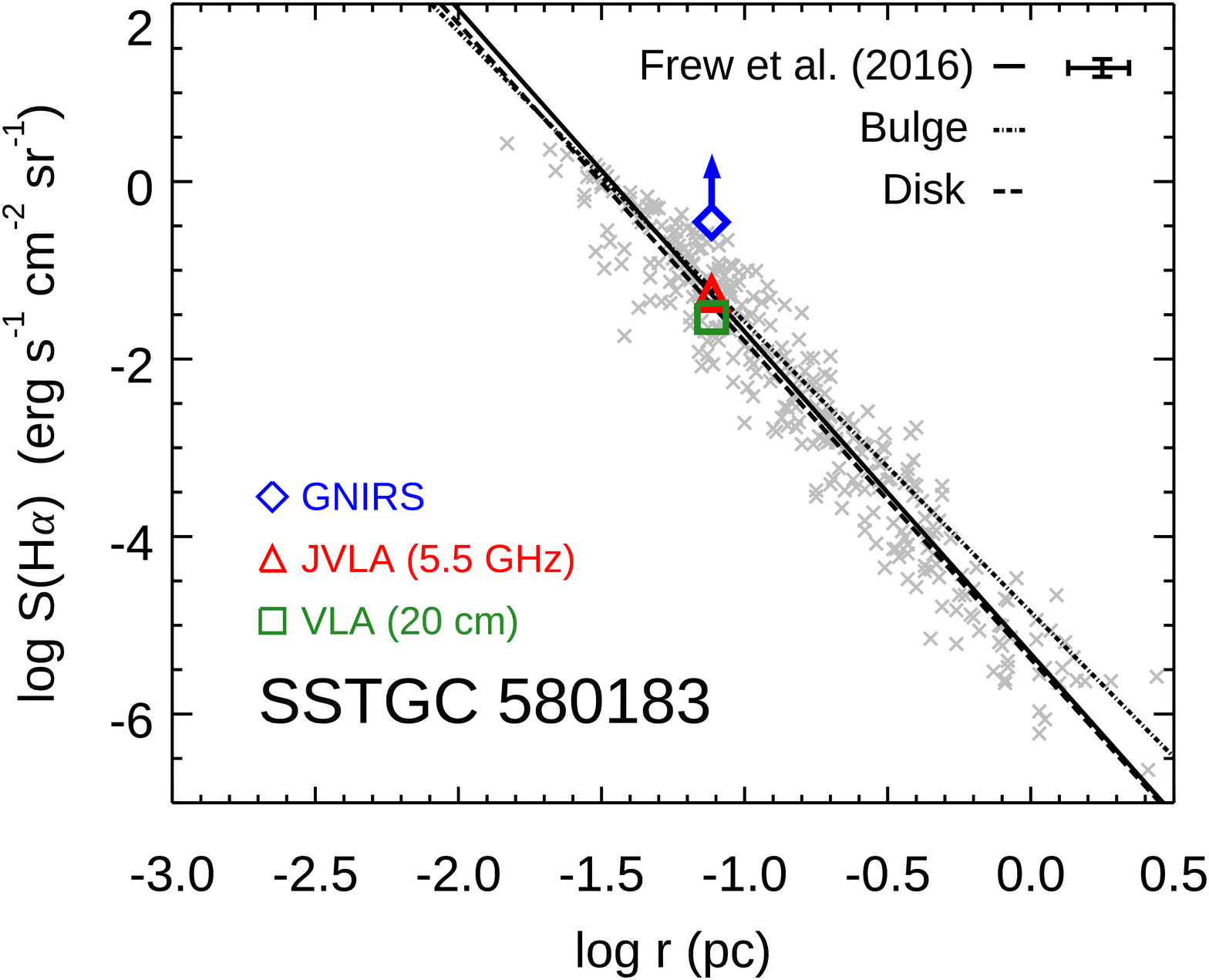}
\includegraphics[scale=0.19]{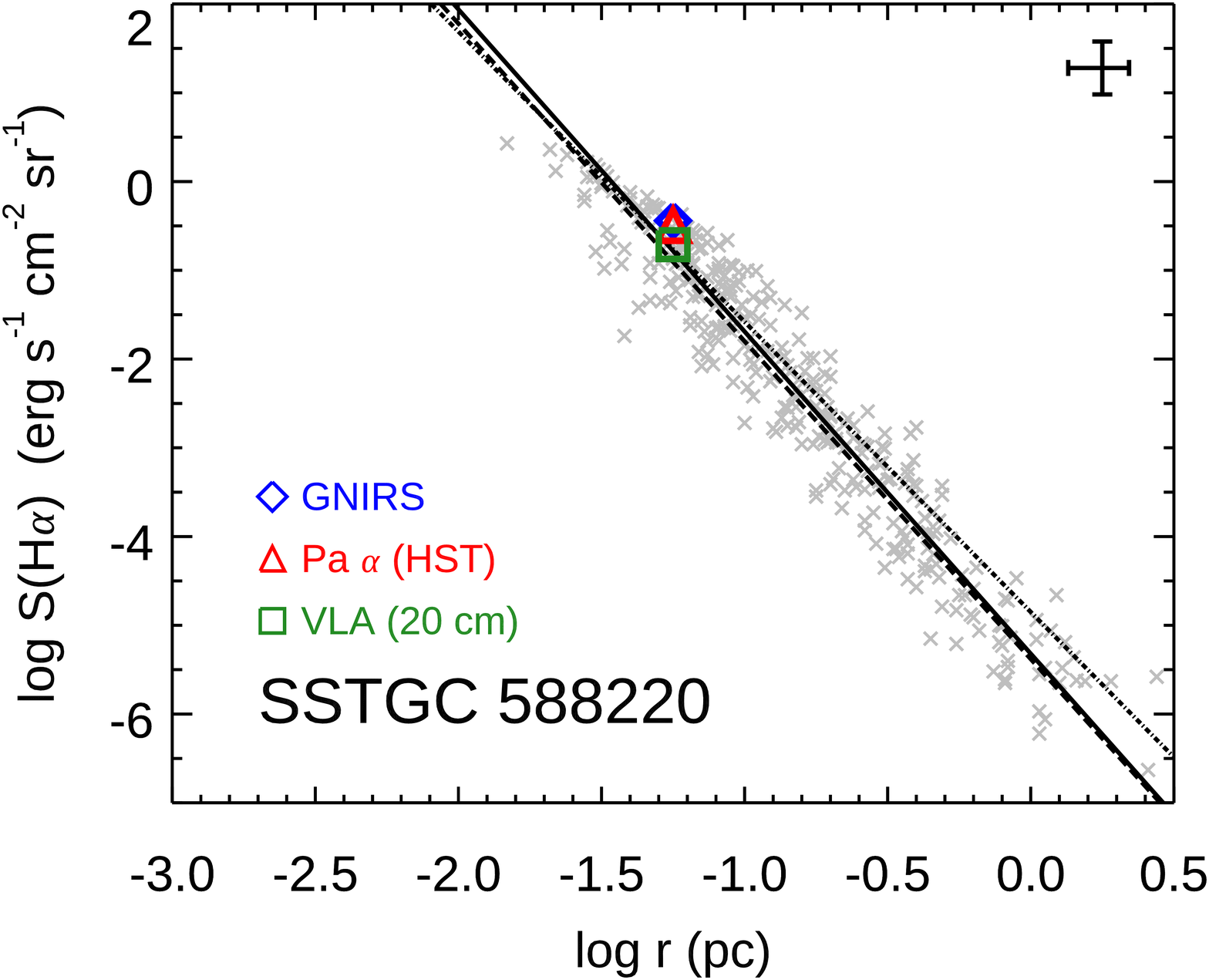}
\caption{Same as Figure~\ref{fig:stanghellini}, but showing comparisons to the radius versus H$\alpha$ surface brightness relation in \citet{frew:16b}. The relations derived from Galactic bulge and disk populations are indicated by the dotted and dashed lines, respectively. The best-fitting power law of their full calibrator sample is shown by a solid line.}
\label{fig:frew}
\end{figure}

Galactic and extra-Galactic PNe exhibit a tight correlation between the radius and surface brightness \citep[e.g.,][see Figures~\ref{fig:stanghellini} and \ref{fig:frew}]{frew:16b,stanghellini:20}, according to which a smaller PN tends to have a higher mean surface brightness. Because other astrophysical nebulae, such as classical nova shells, show noticeable offsets from this relation \citep[e.g.,][]{frew:16a}, its direct comparison can be used not only to confirm the nature of our sources as PNe, but also to constrain a distance range assuming that our sources directly follow the mean PN relation.

For this comparison, we measured the angular size of SSTGC~588220 from the {\it HST}/NICMOS Pa~$\alpha$ image \citep{wang:10}. Since the bright inner rim shows a mild ellipticity (Figure~\ref{fig:image2}), we took the average of the angular diameter measured along the major and minor axes, $2.83\pm0.02\arcsec$. Reassuringly, this size is comparable to the spatial extent of the observed emission line profiles (Figure~\ref{fig:profile}). On the other hand, emission lines from SSTGC~580183 show double peaks along the slit, and the observed full extent is approximately equal to the mean diameter from high-resolution radio images \citep{zhao:20}, $3.86\pm0.04\arcsec$.

For our targets, the H$\alpha$ and H$\beta$ fluxes were computed from the extinction-corrected Br~$\gamma$ line flux, assuming the same Case~B recombination coefficients as in the previous section: $j({\rm H}\alpha)/j({\rm Br}\gamma)=103.0$ and $j({\rm H}\beta)/j({\rm Br}\gamma)=36.1$ \citep{storey:95}. Over $5\times10^3 \leq T_e (K) \leq 2\times10^4$ and $10^2 \leq N_e \leq 10^4$, the ranges of these coefficients are $92.2$--$118$ and $30.3$--$43.0$, respectively. Because the GNIRS targets are more extended than the slit width, we estimated the amount of light lost by simulating our observations with the continuum-subtracted {\it HST} Pa~$\alpha$ images \citep{wang:10}. From this, we found that the $0.675\arcsec$-wide slit collects approximately $41\%\pm3\%$ of the total light from SSTGC~588220. Similarly, we used the $5.5$~GHz map \citep{zhao:20} for SSTGC~580183 to compute the surface area covered by the slit, $18\%\pm4\%$, assuming a $20\%$ uncertainty in this measurement.

To compute uncertainties in the surface brightness estimates, we considered various sources of systematic errors, such as those from the angular diameter measurements, light loss correction, and Case B recombination coefficients, as presented above. In addition, since the H$\alpha$ and H$\beta$ line fluxes depend on which of the three extinction curves included in this study is taken, we explicitly included it as a source of systematic uncertainties. We also assumed a $20\%$ uncertainty in the absolute flux calibration. The final uncertainties were computed by adding in quadrature the above systematic errors and the flux measurement uncertainties.

Figure~\ref{fig:stanghellini} shows the radius versus H$\beta$ surface brightness relation of the Galactic PNe in \citet{stanghellini:20}, which is based on accurate distances to the PNe central stars from {\it Gaia} DR2 \citep{gaia:18}. A comparison to our GNIRS observation for SSTGC~580183 and SSTGC~588220 is shown in the top and bottom panels, assuming that they are located in the GC at a distance of $d_\odot = 8.2$~kpc \citep{gravity:19,reid:19}. In both panels, the average H$\beta$ surface brightness estimated from the GNIRS Br~$\gamma$ flux is shown by blue open diamonds, assuming an average extinction toward each object (Table~\ref{tab:tab2}). Since the lower limit on the extinction correction is available for SSTGC~580183, its $3\sigma$ lower limit is marked. The vertical error bars indicate the quadrature sum of the uncertainties as described above. The horizontal error bars show an approximate range of the physical radius, assuming that they are located in the Galactic bulge at $d_\odot = 8.2\pm2.0$~kpc (see below for independent derivations of the distances).

The gray crosses in Figure~\ref{fig:stanghellini} indicate the individual Galactic PNe in \citet{stanghellini:20}. There is a significant scatter of points below the relation at smaller radii, which is attributed to the low ionized-masses ($<0.01\ M_\odot$) of these objects among other possible causes \citep[see discussion in][]{stanghellini:20}. Accordingly, if our candidate PNe do not follow the mean size versus surface brightness relation, but are outliers with systematically smaller radii, the inferred distances would become significantly larger by an order of magnitude. However, the ionized-masses of SSTGC~580183 and SSTGC~588220 based on radio fluxes ($\sim0.4$--$3\ M_\odot$ assuming $4$--$8$~kpc distances) are larger than the upper limit for their outliers, rejecting hypothesis. The solid line shows the best-fitting relation to the data in \citet{stanghellini:20} without the low ionized-mass sample.

In addition, Figure~\ref{fig:frew} shows comparisons to the size versus H$\alpha$ surface brightness relation of the Galactic and extra-Galactic PNe of \citet{frew:16b}. The gray crosses indicate the Galactic and extra-Galactic `calibrator' sample in their study, which do not show the same large scatter at small radii as that in Figure~\ref{fig:stanghellini}. The black solid line indicates their best-fitting relation to all sample PNe. The dotted and dashed lines show the observed relations for bulge and disk PNe in the Milky Way, respectively, which are not significantly different from each other and from the mean relation.

In all cases, the GNIRS observations are in good agreement with the PN relations if the sources are put at the GC distance. The same conclusion can be drawn using the Pa~$\alpha$ flux measurement from {\it HST}/NICMOS for SSTGC~588220 (red open triangles in the bottom panels of Figures~\ref{fig:stanghellini} and \ref{fig:frew}). As an independent check on these results, we also employed radio continuum fluxes from VLA 20~cm \citep{yusefzadeh:04} and JVLA $5.5$~GHz observations \citep{zhao:20} to infer their H$\alpha$ and H$\beta$ surface brightnesses, which are essentially free of extinction errors. Moreover, the radio images encompass the whole structure of each target and therefore do not need slit-loss corrections. We employed the relation between the hydrogen recombination line flux and the free-free emission in \citet{scoville:03} and assumed the same set of recombination line Case~B emissivities in the previous section. The surface brightness estimates derived from radio fluxes are marked by green boxes (VLA 20~cm) and red triangles (JVLA $5.5$~GHz; top panels) in Figures~\ref{fig:stanghellini} and \ref{fig:frew}, which show good agreement with both the \citet{frew:16b} and \citet{stanghellini:20} relations.

\input{tab3.tex}

Assuming that our candidate PNe follow the surface brightness versus radius relations in the above two studies, we proceed to constrain the range of distances to each object by directly comparing its measured angular size to the inferred physical size for the estimated average surface brightness. The heliocentric distances computed in this way are summarized in Table~\ref{tab:tab3}. The first two rows show distance estimates using the Br~$\gamma$ fluxes measured with GNIRS in this study, based on \citet{frew:16b} and \citet{stanghellini:20}, respectively. For SSTGC~580183, upper distance limits are shown from the lower limit in the foreground extinction. The uncertainties include those propagated from the surface brightness measurements and the fitting coefficients in each of the relations. Within the uncertainties, both PN relations give consistent distance estimates with each other for each PN candidate.

As shown in Table~\ref{tab:tab3}, the distances to SSTGC~588220 derived from the near-IR line fluxes are slightly smaller than those estimated using radio flux measurements, although they are in mutual agreement within the estimated uncertainties. On the other hand, the $3\sigma$ upper distance limits for SSTGC~580183 seem too small compared to those derived from radio fluxes. This implies that our estimated foreground extinction toward SSTGC~580183 may be too large, or that there may be systematic errors in the adopted extinction curves. The average distance to SSTGC~580183 in Table~\ref{tab:tab3} indicates a mean distance from radio observations; an average distance to SSTGC~588220 is derived from near-IR and radio observations. These average distances, as well as the high foreground extinction, indicate that both targets are likely located in the central region of the Milky Way.

\section{Nebular Abundances}\label{sec:cloudy}

\subsection{Cloudy Models}

\input{tab4.tex}

Estimates were made of the abundances of the two candidate PNe using Cloudy {\tt 17.02} \citep{ferland:17}. The input parameters to the Cloudy models are summarized in Table~\ref{tab:tab4}. These parameters were estimated by computing a large number of models with varying values of the effective temperature ($T_{\rm eff}$) of the exciting star \citep[here $\log{g}=6$ white dwarf models at solar abundance taken from][]{rauch:03}, the hydrogen nucleus density $N_H$ (the electron density $N_e$ being variable with depth in the models depending on the local ionization of the multielectron elements), the filling factor $f$ \citep[see][for equations relating $N_H$ and $f$]{simpson:18}, and the abundances of helium and the heavy elements (see below). The total ionizing luminosities $Q$(H) (number of photons s$^{-1}$) were estimated from the radio fluxes measured by \citet{zhao:20} for SSTGC~580183 and by \citet{yusefzadeh:04} for SSTGC~588220, with the assumption of a rounded GC distance of $d_\odot = 8$~kpc, $T_e = 10^4$~K, and the use of Equation~(1) of \citet{simpson:90}. The inner radii were measured from Figures~\ref{fig:image} and \ref{fig:image2}, and were scaled at $d_\odot = 8$~kpc. The final models presented in Tables~\ref{tab:tab5} and \ref{tab:tab6} were selected as those that best fit the measured, extinction-corrected line flux ratios, with an emphasis on the lines with the highest fractional ionization if multiple ionization states were observed. We note that, since all model comparisons are made using line flux ratios, the results for the abundances estimated from the models will not change if the estimated distances are slightly different from $d_\odot = 8$~kpc.

In this modeling effort, we included various mid-IR lines measured with {\it Spitzer} IRS, in addition to the hydrogen and helium lines observed with GNIRS. We utilized the line intensity measurements from \citet{simpson:18}, which were based on the spectral extraction tool {\tt CUBISM} \citep{smith:07} with point-source flux calibration. All of the lines were measured by fitting Gaussians to the spectra. Some lines were remeasured in this study using the IRS analysis program {\tt SMART-IDEA} \citep{higdon:04} for better accuracy. These include [\ion{S}{4}] 10.51, [\ion{Ar}{5}] 13.10, [\ion{Mg}{5}] 13.52, [\ion{Fe}{6}] 14.77, [\ion{P}{3}] 17.88, [\ion{Fe}{2}] 17.94, [\ion{Ar}{3}] 21.83, [\ion{Ne}{5}] 24.32, and [\ion{Fe}{3}] 22.93~\micron. As described in \S~\ref{sec:observation}, background subtraction was performed on a line-by-line basis by taking the average of the high-resolution (IRS {\tt SH} and {\tt LH} modules) fluxes at four nearby positions, while for the low-resolution modules ({\tt SL} and {\tt LL}), background positions were measured from slightly distant positions along the same slits.

\input{tab5.tex}

In Table~\ref{tab:tab5}, the extinction-corrected line fluxes are expressed as line ratios with respect to the hydrogen recombination lines, where \ion{H}{1}~7--4 (Br~$\gamma$) was used for those observed with GNIRS, and \ion{H}{1}~7--6 (12.37~\micron) for the IRS observations. The division into two wavelength intervals was necessary to minimize systematic errors from the different slit sizes -- the GNIRS observations were made with a relatively narrow slit (slit models), while the longer-wavelength IRS slits were large enough (minimum slit width $3.6\arcsec$) to include the whole source (whole nebula models). We corrected the observed near-IR line fluxes for extinction using $A_K$ estimates derived from the \citet{fritz:11} extinction curve (Table~\ref{tab:tab2}). The same extinction curve was used for the mid-IR emission lines up to the peak of the $9.7\ \mu$m silicate feature. At longer wavelengths, however, we employed the \citet{chiar:06} extinction curve because it extends to a longer wavelength \citep[$\sim 35$~\micron\ instead of the 27~\micron\ of][]{fritz:11}. Also, it is based on a source that should have absorption only (the Quintuplet cluster star GCS~3) and not on a source for which one must compensate for dust emission (Sgr~A) in the extinction curve estimate at the longest wavelengths \citep{fritz:11}. The uncertainties in the background-subtracted line intensities include uncertainties from the flux measurements and the background subtraction, which are added in quadrature.

For the reasons described above, we performed separate Cloudy runs to account for the slit models and the whole-nebula models. In the former, computed by integrating the line volume emissivities over a simulated long slit, the near-IR lines Br~$\gamma$, \ion{He}{1} 2.059~\micron, and \ion{He}{2} 2.189~\micron\ lines were used to produce estimates of $T_{\rm eff}$ and the He/H ratio.\footnote{The other $K$-band \ion{He}{1} lines are listed here by their most important components, but are actually blends with lines that are distinguished in Cloudy. We are unable to separate them owing to the moderate spectral resolution of our data.}  In the whole-nebula models, the mid-IR lines, computed by integrating over the whole volume, were used to produce estimates of $N_{\rm H}$, the abundances of the heavy elements, and $f$ for SSTGC~588220. Aside from these details of the integration, all the model parameters (density, temperature, abundances, etc.) were identical for both slit and whole-nebula models. The real objects, of course, do not have constant density, but we do not have enough data to attempt to model the effects of density variations.

We used Cloudy with its `optimize' commands, whereby, in each call, a given parameter or  set of parameters is allowed to vary with the best solution based on the deviation of the model predictions for a set of lines with the observed fluxes. Because of the large number of input parameters, relative to the limited observational information, a uniform density and a constant filling factor were assumed in these models. The elemental abundances were fit by hand by comparing the model output with the observed line flux ratios. The filling factor was computed by optimize command for SSTGC~588220, but this was not successful for SSTGC~580183. Instead, models with a variety of filling factors were computed and the median model with $f = 0.10$ was selected.

Elemental abundances in the models, other than those derived in the modeling, were taken from Cloudy's standard set for PNe, which we revised to use the more recent observations of C/O and N/O ratios of \citet{stanghellini:18}. Whereas the Cloudy standard PN abundances are greatly enriched in carbon (C/O $> 1$), when we computed the medians of \citet{stanghellini:18} tabulated logarithmic carbon, nitrogen, and oxygen abundance differences, we found ratios C/O $=0.767$ and N/O $=0.372$. The latter ratio is close to the Cloudy standard PN abundance ratio and is substantially higher than typical \ion{H}{2} region abundance ratios \citep[e.g.,][]{simpson:18}.
For silicon, we used the GC Si/O ratio of 0.035 from \citet{simpson:18}. As we varied the oxygen abundance in the models, we also varied the C, N, and Si abundances in accordance with these ratios. Since we observed neon and argon lines directly, we did not need to use the \citet{stanghellini:18} ratios. The graphite and silicate grain opacities were included in the calculation.

The criteria for an acceptable fit and the steps in the modeling process were as follows:

(i) The \ion{He}{2} 2.189~\micron/\ion{He}{1} 2.059~\micron\ line ratio, which was used to estimate the $T_{\rm eff}$ of the PN central star, should be equal to the observed ratio. Iteration was needed, since the exact $T_{\rm eff}$ is somewhat dependent on $N_{\rm H}$ and $f$ as they both affect the ionization parameter $U$ \citep[][their Equation~2]{simpson:18}.

(ii) Density-sensitive line pairs (mainly [\ion{S}{3}] 18.7~\micron\ and 33.5~\micron, but also [\ion{Ne}{5}] 14.3~\micron\ and 24.3~\micron\ for SSTGC~588220) should agree with the observed line ratios; the [\ion{Ar}{3}] 8.99~\micron\ and 21.8~\micron\ line pair in SSTGC~580183 was not used because of the large uncertainties. As shown in the high-resolution images (bottom panel of Figures~\ref{fig:image} and \ref{fig:image2}), the PNe may not have a constant density. Indeed, it was not possible to find constant-density models that produced line ratios that agreed with all line pairs. For this reason, less weight was given to the [\ion{S}{3}] line pair in SSTGC~588220, which has larger uncertainties than the [\ion{Ne}{5}] line pair. 

(iii) The line ratios with respect to hydrogen should agree with the observed ratios. To ensure this, the relative abundances of helium and the heavy elements were adjusted in each iteration. In the models, the abundances of carbon, nitrogen, and silicon, all unobserved in our candidate PNe, were scaled to the abundance of oxygen as discussed above. This resulted in additional iterations as changing the abundances modifies the cooling function and hence the electron temperature, which in turn affects all computed fluxes and flux ratios.

\subsection{Results}

\input{tab6.tex}

The elemental abundances in the best-fitting Cloudy models are given in Table~\ref{tab:tab6}. Although we ran models assuming constant $N_{\rm H}$ and $f$, the above modeling exercise provides useful and reliable information on the central star $T_{\rm eff}$ and, in particular, on the abundances of helium, neon, and sulfur, all of which were observed in lines from more than one ionization stage. Of lesser reliability are the elements observed from only a single ionization stage, although some have, at least, the dominant ionization stage observed. The reliability can be estimated from the sizes of the errors.

Uncertainties in the background subtraction propagate into significant uncertainties in the final abundances for most heavy elements. This is because abundances are tabulated with respect to hydrogen, where the strongest hydrogen line in the IRS wavelength is the H~7--6 line at 12.37~\micron, which is sensitive to background subtraction. Whereas te highest-excitation lines have little or no presence in the background spectra, the low-excitation lines have substantial background emission that needs to be removed. In particular, background subtraction removes 27\%\ of the measured target flux of the H~7--6 line for SSTGC~580183, and 40\%\ of the measured target flux for SSTGC~588220 (see Figure~\ref{fig:irs2}); the final S/N are $18$ and $10$ for SSTGC~580183 and SSTGC~588220, respectively (Table~\ref{tab:tab5}). The other singly ionized lines ([\ion{Ne}{2}], [\ion{Ar}{2}], and [\ion{Fe}{2}]) have even more flux subtracted (so much so for [\ion{Fe}{2}] that this line was dropped from the analysis), but these elements have doubly ionized lines in the IRS spectra, which make our abundance estimates more secure. For these calculations, the measurement uncertainty includes the uncertainty in the background subtraction, computed as the sum, in quadrature, of the measured uncertainties in the line fluxes for target and background positions. It does not include any calibration uncertainty or estimation of any nonlinear departures from a smoothly sloping background.

The abundance uncertainties in Table~\ref{tab:tab6} are the sum, in quadrature, of the measurement uncertainty, the uncertainty due to a possible error in the extinction of $\Delta \tau_{9.6}=0.5$, and an estimate of the model error expressed as the uncertainty in the ionization correction factor ($ICF$). An $ICF$ is defined here as the ratio of the total abundance of an element to the abundance of the observed ionization stage as computed by the model. The $ICF$ uncertainty is estimated to be the value of 0.05 divided by the fractional ionization of the observed ion for elements where only one ionization state was observed or it is estimated to equal 0.25 times the fractional ionization of the unobserved ionization states when multiple ionization states were observed. Which of these errors dominates depends on the line: the measurement error is important for the two blended lines, [\ion{P}{3}] 17.88~\micron\ and [\ion{Fe}{2}] 17.94~\micron\ (the latter having such a large uncertainty that it was not further included); the extinction error is mainly important for the lines with wavelengths near 10\ \micron\ and 18~\micron\ (this uncertainty is large because the silicate extinction curve is not that well known); and the $ICF$ error dominates for lines arising from ions with very low fractional ionization, such as O$^{3+}$ and Mg$^{4+}$. Finally, the uncertainty for O/H in SSTGC~588220 was increased because the line was saturated in the high-resolution IRS module ({\tt LH}) and the flux had to be taken from the low-resolution ({\tt LL1}) observation.

\subsubsection{SSTGC~588220 (G0.0967-0.0511)}

The fluxes predicted by the Cloudy models provide reasonably good agreement with the observed fluxes. However, in spite of the overall agreement, notable discrepancies are found in some lines. For instance, the best-fitting model overpredicts the fluxes from [\ion{Ar}{5}] 13.1~\micron\ and [\ion{Fe}{6}] 14.7~\micron, while it underestimated the \ion{He}{1} 1.083~\micron\ flux. The overestimated high-excitation lines are probably a result of the source not being spherically symmetric and having large density fluctuations. On the other hand, the \ion{He}{1} 1.083~\micron\ line is more sensitive to the amount of extinction than the longer-wavelength He lines, and its emissivity is more dependent on the details of the electron density distribution than the emissivity of the \ion{He}{1} 2.059~\micron\ line \citep[see][]{porter:12,porter:13}.

\subsubsection{SSTGC~580183 (G359.9627-0.1202)}

The agreement with the best-fitting model is less satisfactory in the case of SSTGC~580183, most likely due to the PN having components of multiple densities and filling factors, rather than having constant values as assumed in the current analysis. In particular, the optimize command in Cloudy initially produced a very low filling factor ($f\approx0.03$) in the models, driven by the absence of high-ionization lines such as [\ion{Ne}{5}], [\ion{Ar}{5}], and [\ion{Mg}{5}] in the {\it Spitzer} mid-IR spectra, except [\ion{O}{4}] 26~\micron. The models described in Tables~\ref{tab:tab4} and \ref{tab:tab5} were produced with $f=0.1$ (the model parameters do not change appreciably with changing $f$) and an inner radius of $0.025$~pc (estimated from the ring size in Figure~\ref{fig:image}, bottom). The uncertainties in these tables do not include any uncertainty for choice of model, but an additional $\sim 10$\%\ should be added to the abundance uncertainties in the tables to account for variations in the possible model parameters. 

Notably, the oxygen abundance from the best-fitting model is supersolar (Table~\ref{tab:tab6}). With a low $f$, the fraction of elements in the highest-ionization states (like O$^{3+}$) is very small, and the oxygen abundance (O/H) becomes substantially higher than solar. We also searched for models of more compact PNe with lower $f$ and smaller inner radii that produce a higher ionization fraction of oxygen. However, we found that none of these models could produce as strong [\ion{O}{4}] emission as is observed unless the total oxygen abundance is substantially higher than solar. In addition, because the $T_{\rm eff}$ of the central star can be reliably determined from the \ion{He}{2} 2.189~\micron/\ion{He}{1} 2.059~\micron\ line ratio, and both He$^{++}$ and O$^{3+}$ require $> 54$~eV photon energies for ionization to that level, the higher oxygen abundance is less affected by the choice of spectral energy distribution of the exciting star.

Along with oxygen, the helium abundance is also significantly higher than in SSTGC~588220. Because the emission from heavy element ions, especially oxygen, is the primary coolant for the gas, increasing their abundances results in lower electron temperatures. The lower temperatures, in turn, increase the emissivities of the hydrogen recombination lines much more than they change the emissivities of the \ion{He}{1} 2.059~\micron\ line \citep[see][]{storey:95,porter:12,porter:13}. The consequence is that the abundance of helium with respect to hydrogen must be increased above solar to match the observed line fluxes.

\subsection{[\ion{Kr}{3}] 2.199~\micron}

In both candidate PNe, we detected the $s$-process noble gas krypton (Kr) at 2.199~\micron, which was first identified in PNe by \citet{dinerstein:01}. From the line flux measurements, we computed Kr$^{++}$/H$^+$ ratios of $(3.98\pm1.80)\times10^{-9}$ and $(1.36\pm 0.13)\times10^{-9}$ for SSTGC~580183 and SSTGC~588220, respectively, based on the Br~$\gamma$ emissivities of \citet{storey:95}, the effective collisional strengths for Kr$^{++}$ of \citet{schoning:97}, and the transition probabilities for Kr$^{++}$ of \citet{eser:19}. Because [\ion{Kr}{3}] 2.199~\micron\ is not modeled in the current version of Cloudy, we adopted the analytical relation for the $ICF$ from \citet{sterling:15} between the Kr$^{++}$/Kr and S$^{++}$/S ratios, which show the tightest fit among other ionic ratios (such as Ar$^{++}$/Ar). In addition, our S$^{++}$/S ratio is more reliable than Ar$^{++}$/Ar, because the S$^{++}$ and S$^{3+}$ measurements were made using higher-resolution modules in IRS with less extinction. The Kr/H abundance ratios are included in Table~\ref{tab:tab6}. The uncertainties do not include the uncertainty in the $ICF$.

\subsection{Significance}\label{sec:sig}

Krypton is significantly enriched in both of our targets ([Kr/Fe] $=1.4\pm0.3$ and $0.4\pm0.2$ for SSTGC~580183 and SSTGC~588220, respectively)\footnote{Here, we used conventional notation, [X/Y] $\equiv {\log{(N_X / N_Y)}}_*- {\log{(N_X / N_Y)}}_\odot$, where $\log{N}$ indicates the abundances of specific elements.}, indicating that this $s$-process element has been overproduced, possibly during the late asymptotic giant branch (AGB) evolution. This is in agreement with other lines of evidence supporting their status as PNe (morphology, high-excitation lines, and a match to the PN size-surface brightness relations) presented in this paper.

The abundance patterns of our targets are generally consistent with those in the GC region. Stars in this region have a bimodal metallicity distribution with peaks at [Fe/H]$\sim-0.5$ and $\sim+0.3$, and exhibit an elevated $\alpha$-element abundance in the metal-poor component, while the metal-rich counterpart shows near-solar abundance patterns \citep[e.g.,][]{schultheis:20}. Reassuringly, our derived metallicities, [Fe/H]$\approx-0.5\pm0.2$ for SSTGC~580183 and [Fe/H]$\approx+0.1\pm0.2$ for SSTGC~588220, coincide with these metallicity peaks. Regarding $\alpha$-element abundances, SSTGC~588220 shows abundance patterns ([O/Fe]$=0.2$, [Mg/Fe]$=0.4$) that are similar to the mean trend of the APOGEE GC sample \citep{schultheis:20} within $\Delta$[$\alpha$/Fe] $\la 0.1$~dex. On the other hand, the oxygen abundance of SSTGC~580183 is quite uncertain, although the large sulfur abundance ([S/Fe]$=1.0\pm0.3$) seems to suggest enhanced light element abundances of this candidate PN.

Our radial velocity measurements are weighted mean values from all observed near-IR lines.\footnote{More precisely, our radial velocity measurements are based on nebular emission lines, and therefore may not be the same as those of the central star of a PN
\citep[e.g.,][]{lorenzo:21}. Since the expansion speed of typical PNe is in the order of a few tens of kilometers per second, our reported values can differ from the radial velocity of a central star by this amount.} A high positive radial velocity of SSTGC~580183 ($v_r^{\rm LSR}\sim+70$~km~s$^{-1}$) in the fourth Galactic quadrant and an extreme negative radial velocity of SSTGC~588220 ($v_r^{\rm LSR}=-150$~km~s$^{-1}$) in the first quadrant are at odds with the sense of rotation of the nuclear stellar disk (a positive correlation between $v_r$ and $l$), but there is a large scatter in the $v_r$ vs.\ $l$ diagram of stars in the nuclear bulge \citep{schonrich:15,schultheis:20}.

SSTGC~580183 is helium-rich ($N$(He)/$N$(H)$=0.12$), while SSTGC~588220 has a nearly solar helium abundance. According to the PN classification scheme in \citet{peimbert:78}, the helium abundance of SSTGC~580183 is at the border between helium- and nitrogen-rich PNe (Type~I) and other types with normal helium abundances. Since Type~I PNe are considered descendants of a massive progenitor star ($> 2\ M_\odot$) by their super-helium and super-nitrogen abundances \citep[see also][]{stanghellini:10}, it is conceivable that the progenitor of SSTGC~580183 is likely to be as massive as $\sim2\ M_\odot$. Such massive stars are not currently present in old stellar populations of the GC, and therefore other formation channels such as binary interactions \citep[e.g.,][]{minniti:19} may be a viable path to the object's formation. Conversely, SSTGC~580183 may simply be a product of recent star formation in the CMZ \citep{an:11,longmore:13}. Below, we further discuss the possible formation mechanisms of these candidate PNe on the basis of the low detection rate of PNe in the CMZ.

\section{Discussion and Summary}\label{sec:summary}

In this paper, we have presented medium-resolution near-IR spectra taken with Gemini/GNIRS of two candidate PNe that were serendipitously found in {\it Spitzer}/IRS spectra in close lines of sight to Sgr~A$^*$. Besides showing strong emission lines in the mid-IR from several high-excitation ions, their appearance in high-resolution images supports their status as PNe. Moreover, our Gemini/GNIRS near-IR spectra reveal strong emission from doubly ionized krypton in both targets, indicating overproduction of $s$-process elements in the PN envelope. Their membership in the nuclear stellar disk seems feasible from the proximity to the GC on the sky and distances based on a comparison to PN size-surface brightness relations, and is favored by our joint analysis of near- and mid-IR spectra, over membership in other Galactic structural components. Our Cloudy modeling assumes uniform densities and filling factors of these targets, but future observations with high spatial resolution mapping in various near- and mid-IR lines can potentially reveal high-density or even high-excitation structures within the overall morphology of these sources. 

The expected number of PNe in the nuclear bulge may be estimated from a comparison with other Galactic stellar components. The total stellar mass of the nuclear bulge \citep[$\sim1.4\times10^9\ M_\odot$;][]{launhardt:02} is approximately $10$ times smaller than that of the classical bulge. If we assume that most stars in the nuclear bulge are as old as those in the classical bulge ($\ga10$~Gyr), and simply scale the result obtained from the population synthesis models in \citet{moe:06} by stellar mass, there should be a significant number of PNe ($\sim10^3$) in this region, although this estimate is uncertain by a factor of two or more.

At the other extreme, a strict lower limit on the expected number of PNe may be estimated from a direct comparison with the number of PNe in the Galactic halo, even though the chemical properties and ages of stars in the nuclear disk are significantly different from those in the halo, and may affect the production rate of PNe. On the theoretical side, if we simply scale the expected number for the stellar halo in \citet{moe:06} by mass, the expected number of PNe becomes $\sim50$, but this estimate is quite uncertain. The large reduction in the expected number of PNe is due to the older ages of the stellar halo (a relatively shorter dissipation timescale of stellar envelopes) compared to those of the bulge populations in their models. On observational grounds, there are eight PNe known in the Galactic globular clusters \citep{jacoby:97,minniti:19}. Therefore, one can naively expect $\sim10^3$ PNe, since the nuclear bulge is $\sim100$ times more massive than the entire globular cluster system in the Milky Way. In this regard, the observed paucity of PNe in the nuclear disk raises a more severe problem than that in other Galactic components with predominantly old stellar populations.

To explain the observed scarcity of PNe in the nuclear bulge, the following scenarios can be speculated. Firstly, this may be due to extreme foreground extinction toward the GC. The average extinction toward the CMZ is $A_V \sim 30$~mag, but the patchy extinction can be as high as $A_V > 60$~mag \citep{schultheis:09}. However, the two PN candidates presented in this work have foreground extinctions that are not exceptionally smaller than those in other fields. Unless PNe in the nuclear disk are preferentially enshrouded by dusty clouds, the large extinction may not be a dominant factor for the low detection rate of PNe. Indeed, extended sources with PN-like morphologies are rare in essentially extinction-free, 20~cm radio continuum images \citep{yusefzadeh:04}, which cover the entire CMZ. Such objects are also unusual in the {\it HST}/NICMOS Pa~$\alpha$ images \citep{wang:10}, except SSTGC~588220, in spite of a large sky coverage of the survey -- a central $\sim39\times15$ arcmin$^2$ region of the CMZ, including Sgr~A$^*$.

Second, there may exist an observational bias against PNe with large envelope sizes. The two PNe presented in this paper have the most compact sizes (implying early stages of PN evolution) as can be seen from the size-surface brightness distributions of other Galactic and extra-Galactic PNe (see Figures~\ref{fig:stanghellini} and \ref{fig:frew}). Their compactness has even made them appear as pointlike sources in the original {\it Spitzer}/IRAC catalog \citep{ramirez:08}. As a consequence, there could be more PNe in lines of sight toward the CMZ that are more difficult to detect due to extended envelopes with lower surface brightness in the high source density, highly dust-obscured region.

Third, if the observed lack of PNe in the GC have astrophysical origins, this may indicate that the PN formation channel such as binary interactions \citep[e.g.,][]{minniti:19} is overpowered by suppression mechanisms that can reduce the envelope mass of AGB stars. Some stars can also skip the AGB phase and directly evolve into hot subdwarfs, if the envelope mass is not high enough. These AGB manqu\'{e} stars can be formed from helium-rich progenitor stars by enhanced mass loss during the red giant branch evolution \citep[][]{bressan:12}. Such helium-rich stars are expected to be more common in a deep potential well, such as in the GC, where chemical enrichment runs relatively fast.

Lastly, while PNe are strikingly rare in the nuclear bulge, the number of observed PNe is not greatly different from what one would expect from ongoing star formation in the CMZ, with a rate of $\sim0.1\ M_\odot\ yr^{-1}$ \citep[e.g.,][]{an:11,longmore:13}. Since the PN lifetime is approximately $\Delta t\sim10^4$~years, the total mass of stars formed at any time interval $\Delta t$ is $\sim10^3\ M_\odot$, if the star formation has proceeded constantly over the past few billion years. If stars as massive as $2\ M_\odot$ can evolve into a PN, for the reasons described above, the total mass of PN progenitors would be $\sim20\ M_\odot$ at any time assuming a standard mass function ($\propto M^{-2.3}$). Therefore, the expected number of PNe would become $\sim5$--$10$, if the average progenitor mass is $2$--$4\ M_\odot$ in the nuclear bulge. This number is significantly smaller than the above estimates based on \citet{moe:06}, which assume that stellar populations in the nuclear bulge are composed entirely of passively evolving old stars. Interestingly, \citet{simpson:18} found half a dozen more candidates with exceptionally highly-excited lines from neon and/or oxygen, in addition to the two sources included in this study, which serve as good PN candidates in this region. Future high-resolution narrowband imaging or mid-IR spectroscopic surveys will help us hunt for additional PNe in the nuclear bulge.

\acknowledgments

We thank the anonymous referee for detailed comments, which helped to improve the content and presentation of the paper. We thank Kris Sellgren and Harriet Dinerstein for many helpful and interesting discussions over the years. We thank Thomas Geballe for his technical assistance before and during the observing runs. J.H.\ and D.A.\ acknowledge support provided by the National Research Foundation of Korea (NRF) grants funded by the Korean government (MSIT) (No. 2017R1A5A1070354, 2018R1D1A1A02085433, 2021R1A1C1004117).

The Gemini data (Gemini program GN-2016A-Q-42, NOAO prop.~ID 2016A-0431), acquired through the Gemini Observatory Archive at the NSF's NOIRLab and processed using the Gemini IRAF package, are based on observations obtained at the international Gemini Observatory, a program of NSF's NOIRLab, which is managed by the Association of Universities for Research in Astronomy under a cooperative agreement with the National Science Foundation on behalf of the Gemini Observatory Partnership: the National Science Foundation (United States), the National Research Council (Canada), Agencia Nacional de Investigaci\'{o}n y Desarrollo (Chile), Ministerio de Ciencia, Tecnolog\'{i}a e Innovaci\'{o}n (Argentina), Minist\'{e}rio da Ci\^{e}ncia, Tecnologia, Inova\c{c}\~{o}es e Comunica\c{c}\~{o}es (Brazil), and Korea Astronomy and Space Science Institute (Republic of Korea). This work was made possible by observations made from the Gemini North telescope, located within the Maunakea Science Reserve and adjacent to the summit of Maunakea. We are grateful for the privilege of observing the universe from a place that is unique in both its astronomical quality and its cultural significance.

{}

\end{document}

%% file: tab1.tex
\begin{deluxetable}{lccc}
\tablecaption{Line Flux Measurements from GNIRS Spectra\label{tab:tab1}}
\tabletypesize{\scriptsize}
\tablehead{
\colhead{} &
\colhead{Wavelength} &
\multicolumn{2}{c}{Flux ($10^{-15}$ erg s$^{-1}$ cm$^{-2}$)} \\
\colhead{Line} &
\colhead{($\mu$m)} &
\colhead{SSTGC~580183} &
\colhead{SSTGC~588220}
}
\startdata
\ion{He}{1}       & $1.0833$ & $<0.256$        & $ 5.311\pm0.078$ \\
Pa~$\beta$        & $1.2822$ & $<0.256$        & $ 6.423\pm0.099$ \\
Br 12--4          & $1.6412$ & $<0.256$        & $ 1.405\pm0.038$ \\
$[$\ion{Fe}{2}$]$ & $1.6440$ & $<0.256$        & $ 1.049\pm0.212$ \\
H$_2$ 1-0 S(2)    & $2.0338$ & $<0.256$        & $ 0.201\pm0.045$ \\
\ion{He}{1}       & $2.0587$ & $6.140\pm0.389$ & $ 10.413\pm0.057$ \\
\ion{He}{1}       & $2.1128$ & $0.387\pm0.020$ & $ 1.178\pm0.051$ \\
\ion{He}{1}       & $2.1137$ & $<0.256$        & $ 0.663\pm0.223$ \\
H$_2$ 1-0 S(1)    & $2.1218$ & $0.145\pm0.059$ & $ 0.738\pm0.045$ \\
\ion{He}{1}       & $2.1615$ & $<0.256$        & $ 0.719\pm0.049$ \\
\ion{He}{1}       & $2.1649$ & $0.626\pm0.041$ & $ 2.339\pm0.201$ \\
Br~$\gamma$       & $2.1661$ & $8.370\pm0.395$ & $31.395\pm0.443$ \\
\ion{He}{2}       & $2.1884$ & $0.252\pm0.022$ & $ 3.143\pm0.044$ \\
$[$\ion{Kr}{3}$]$ & $2.1990$ & $0.249\pm0.112$ & $ 0.403\pm0.037$ \\
\hline
$\langle v_r^{\rm helio} \rangle$ & & $+62\pm2$~km~s$^{-1}$ & $-160\pm9$~km~s$^{-1}$ \\
$\langle v_r^{\rm LSR} \rangle$ & & $+72\pm2$~km~s$^{-1}$ & $-150\pm9$~km~s$^{-1}$ \\
\enddata
\tablecomments{Fluxes with $1\sigma$ uncertainties are shown. Upper limits are indicated as $3\sigma$ detection limits.}
\end{deluxetable}

%% file: tab2.tex
\begin{deluxetable}{lcc}
\tablecaption{Foreground Extinction Estimates\label{tab:tab2}}
\tablehead{
\colhead{Extinction} &
\colhead{SSTGC~580183\tablenotemark{\scriptsize a}} &
\colhead{SSTGC~588220}
}
\startdata
$A_K$ \citep{fritz:11} & $>2.72$ & $1.74\pm0.03$ \\
$A_K$ \citep{chiar:06} & $>3.64$ & $2.33\pm0.04$ \\
$A_K$ \citep{boogert:11} & $>3.95$ & $2.53\pm0.05$ \\
\hline
Weighted average $\langle A_K \rangle$ & $>3.26$ & $2.09\pm0.48$ \\
Weighted average $\langle A_V \rangle$\tablenotemark{\scriptsize b}        & $>29.7$  & $19.0\pm4.4~$ \\
$\langle \tau_{9.6} \rangle$\tablenotemark{\scriptsize c} & $>5.06$  & $3.24\pm0.71$ \\
\hline
$\langle A_K \rangle$ \citep{schultheis:09}\tablenotemark{\scriptsize d} & $3.76\pm0.49$  & $2.90\pm0.17$ \\
$\langle \tau_{9.6} \rangle$ \citep{an:13}\tablenotemark{\scriptsize e}   &  $2.89\pm0.37$  & $3.69\pm0.31$ \\
$\langle \tau_{9.6} \rangle$ \citep{simpson:18}   &  $2.73$  & $2.80$ \\
\enddata
\tablecomments{Uncertainties are the $1\sigma$ standard deviation of each estimated variable.}
\tablenotetext{a}{$3\sigma$ upper limits, unless uncertainties are indicated.}
\tablenotetext{b}{Assuming $A_K/A_V=0.11$ \citep{figer:99}.}
\tablenotetext{c}{Assuming $1.3 \leq \tau_{9.6}/A_K \leq 1.8$ (see text).}
\tablenotetext{d}{Extinction within $2\arcmin$ of the source.}
\tablenotetext{e}{Extinction within $0.9\arcmin$--$1.5\arcmin$ of the source.}
\end{deluxetable}

%% file: tab3.tex
\begin{deluxetable}{lcc}
\tablecaption{Heliocentric Distance Estimates\label{tab:tab3}}
\tablehead{
\colhead{Observations/Relation\tablenotemark{\scriptsize a}} &
\colhead{SSTGC~580183\tablenotemark{\scriptsize b}} &
\colhead{SSTGC~588220}
}
\startdata
Br~$\gamma$/\citeauthor{frew:16b} & $<4.8$~kpc & $6.5\pm1.8$~kpc \\
Br~$\gamma$/\citeauthor{stanghellini:20} & $<5.5$~kpc & $7.4\pm1.5$~kpc \\
Pa~$\alpha$/\citeauthor{frew:16b} & \nodata & $6.7\pm1.8$~kpc \\
Pa~$\alpha$/\citeauthor{stanghellini:20} & \nodata & $7.6\pm1.5$~kpc \\
\hline
20~cm/\citeauthor{frew:16b} & $9.5\pm1.5$~kpc & $7.7\pm1.4$~kpc \\
20~cm/\citeauthor{stanghellini:20} & $9.9\pm2.1$~kpc & $8.6\pm1.9$~kpc \\
5.5~Ghz/\citeauthor{frew:16b} & $8.1\pm1.3$~kpc & \nodata \\
5.5~Ghz/\citeauthor{stanghellini:20} & $8.6\pm1.8$~kpc & \nodata \\
\hline
Average & $9.0\pm1.6$~kpc\tablenotemark{\scriptsize c} & $7.6\pm1.6$~kpc \\
\enddata
\tablecomments{Uncertainties are the $1\sigma$ standard deviation of each estimated variable.}
\tablenotetext{a}{Observations: Wavelengths at which fluxes were measured to derive surface brightness -- Br~$\gamma$ (this study), Pa~$\alpha$ \citep{wang:10}, $20$~cm \citep{yusefzadeh:04}, and $5.5$~GHz \citep{zhao:20}. Relations: Surface brightness versus size relations of PNe in \citet{frew:16b} and \citet{stanghellini:20} used to compute the physical size of a PN.}
\tablenotetext{b}{$3\sigma$ upper limits, unless uncertainties are indicated.}
\tablenotetext{c}{Weighted average distance from radio observations.}
\end{deluxetable}

%% file: tab4.tex
\begin{deluxetable}{lccc}
\tablecaption{Physical Parameters Used in the Cloudy Models\label{tab:tab4}}
\tablehead{
   \colhead{Parameters} &
   \colhead{SSTGC~580183} & 
   \colhead{SSTGC~588220} &
   \colhead{}
}
\tablehead{\colhead{} & \colhead{SSTGC~580183} & \colhead{SSTGC~588220} & \colhead{} }
\startdata
$T_{\rm eff}$ (K)\tablenotemark{\scriptsize a} & $105,947$        & $131,856$ \\
Log Q(H) (s$^{-1}$)                & $46.892         $ & $47.194$ \\
Inner Radius (pc)                  & $0.0250          $ & $0.02235$ \\
Electron Density ($N_e$; cm$^{-3}$)       & $5425 \pm 2300 $ & $1580 \pm 800$ \\
Hydrogen Density ($N_H$; cm$^{-3}$)       & $3606          $ & $3025$ \\
Filling Factor ($f$)                     & $0.1000       $ & $ 0.2861$ \\
\enddata
\tablenotetext{a}{Effective temperature of an ionizing star from stellar atmosphere models at $\log{g}=6$ \citep{rauch:03}.}
\end{deluxetable}

%% file: tab5.tex

\begin{deluxetable*}{lccccccccc}
\tablecaption{Normalized Line Fluxes\label{tab:tab5}}
\tabletypesize{\scriptsize}
\tablehead{
   \colhead{} &
   \colhead{} &
   \multicolumn{4}{c}{SSTGC~580183} &
   \multicolumn{4}{c}{SSTGC~588220} \\
   \cline{3-6}
   \cline{7-10}
   \colhead{Emission} &
   \colhead{Wavelength} &
   \multicolumn{2}{c}{Observation} &
   \multicolumn{2}{c}{Model} &
   \multicolumn{2}{c}{Observation} &
   \multicolumn{2}{c}{Model} \\
   \cline{3-4}
   \cline{5-6}
   \cline{7-8}
   \cline{9-10}
   \colhead{Line} &
   \colhead{($\mu$m)} &
   \colhead{Flux\tablenotemark{\scriptsize a}} &
   \colhead{Uncertainty\tablenotemark{\scriptsize b}} &
   \colhead{Flux\tablenotemark{\scriptsize c}} &
   \colhead{Ion.\tablenotemark{\scriptsize d}} &
   \colhead{Flux\tablenotemark{\scriptsize a}} & 
   \colhead{Uncertainty\tablenotemark{\scriptsize b}} &
   \colhead{Flux\tablenotemark{\scriptsize c}} &
   \colhead{Ion.\tablenotemark{\scriptsize d}}}
\startdata
\multicolumn{10}{c}{Gemini GNIRS (slit models)} \\
\cline{1-10}
Pa $\beta$        & $ 1.2822$ & \nodata   & \nodata  & \nodata   & \nodata  & $ 5.877$  &  $0.015$ & $  5.579$ & $-0.028$ \\
Br 12--4      & $ 1.6412$ & \nodata   & \nodata  & \nodata   & \nodata  & $ 0.168$  &  $0.031$ & $  0.186$ & $-0.028$ \\
Br $\gamma$      & $ 2.1661$ & $  1.000$ & $0.047 $ & $  1.000$ & $-0.037$ & $ 1.000$  &  $0.014$ & $  1.000$ & $-0.028$ \\
He I $2^3P$--$2^3S$ & $ 1.0833$ & \nodata   & \nodata  & \nodata   & \nodata  & $40.445$  &  $0.020$ & $ 8.412$ & $-0.221$ \\
He I $2^1P$--$2^1S$ & $ 2.0587$ & $  0.9846$ & $0.079 $ & $ 0.9840$ & $-0.046$ & $ 0.4003$ & $0.015$ & $ 0.4001$ & $-0.221$ \\
He I $4^3P$--$3^3S$ & $ 2.1128$ & $  0.0532$ & $0.070 $ & $ 0.0388$ & $-0.046$ & $ 0.041$ & $0.046$ & $ 0.016$ & $-0.221$ \\
He I $4^1P$--$3^1S$ & $ 2.1137$ & \nodata  & \nodata  & \nodata  & \nodata  & $ 0.023$  &  $0.336$ & $  0.006$ & $-0.221$ \\
He I $7^3F$--$4^3F$ & $ 2.1615$ & \nodata  & \nodata  & \nodata  & \nodata  & $ 0.023$  &  $0.070$ & $  0.011$ & $-0.221$ \\
He I $7^3G$--$4^3F$ & $ 2.1649$ & \nodata  & \nodata  & \nodata  & \nodata  & $ 0.075$  &  $0.087$ & $  0.017$ & $-0.221$ \\
He II 10--7         & $ 2.1884$ & $ 0.0285$ & $0.101 $ & $0.0285$ & $-1.256$ & $ 0.0966$ & $0.020$ & $ 0.0968$ & $-0.426$ \\
$[$\ion{Fe}{2}$]$   & $ 1.6440$ & \nodata   & \nodata  & \nodata   & \nodata  & $ 0.225$  &  $0.203$ & $ 11.93$ & $-0.825$ \\
$[$\ion{Kr}{3}$]$   & $ 2.1990$ & 0.02745  & 0.437  & \nodata   & $-0.437$  & $ 0.0122$  &  $0.0929$ & \nodata & $-0.534$ \\
\cline{1-10}
\multicolumn{10}{c}{{\it Spitzer} IRS (whole nebula models)} \\
\cline{1-10}
H 7--6              & $12.3719$ & $ 1.000$ & $ 0.055$ & $  1.000$ & $-0.052$ & $ 1.000$  &  $0.099$ & $ 1.000$ & $-0.041$ \\
$[$\ion{O}{4}$]$    & $25.8903$ & $ 63.7$ & $ 0.15$ & $ 63.85$ & $-1.931$ & $330.95$  &  $0.102$ & $ 330.98$ & $-0.869$ \\
$[$\ion{Ne}{2}$]$   & $12.8135$ & $ 41.8$ & $ 0.057$ & $ 49.17$ & $-0.765$ & $ 15.57$  &  $0.111$ & $ 29.94$ & $-1.193$ \\ 
$[$\ion{Ne}{3}$]$   & $15.5551$ & $408.6$ & $ 0.056$ & $346.4$ & $-0.092$ & $413.37$  &  $0.099$ & $ 477.8$ & $-0.110$ \\
$[$\ion{Ne}{5}$]$   & $14.3217$ & \nodata   & \nodata  & \nodata   & $-5.209$ & $160.13$  &  $0.099$ & $ 137.5$ & $-1.669$ \\
$[$\ion{Ne}{5}$]$   & $24.3175$ & \nodata   & \nodata  & \nodata   & $-5.209$ & $113.35$  &  $0.099$ & $ 94.0$ & $-1.669$ \\
$[$\ion{Na}{3}$]$   & $ 7.3169$ & $  5.36$ & $ 0.077$ & $  5.34$ & $-0.158$ & \nodata   &  \nodata & \nodata   & $-0.156$ \\
$[$\ion{Mg}{5}$]$   & $13.5213$ & \nodata   & \nodata  & \nodata   & $-5.966$ & $ 0.654$  &  $0.283$ & $  0.652$ & $-1.772$ \\
$[$\ion{P}{3}$]$    & $17.8850$ & $  0.92$ & $ 0.47$ & $  0.914$ & $-0.448$ & $ 1.716$  &  $0.307$ & $  1.716$ & $-0.411$ \\
$[$\ion{S}{3}$]$    & $18.7130$ & $ 120.3$ & $ 0.057$ & $ 106.64$ & $-0.256$ & $81.82$  &  $0.102$ & $ 92.22$ & $-0.295$ \\
$[$\ion{S}{3}$]$    & $33.4810$ & $ 36.7$ & $ 0.23$ & $  32.55$ & $-0.256$ & $40.02$  &  $0.316$ & $ 34.53$ & $-0.295$ \\
$[$\ion{S}{4}$]$    & $10.5105$ & $ 93.4$ & $ 0.056$ & $ 105.27$ & $-1.052$ & $245.80$  &  $0.099$ & $ 218.1$ & $-0.670$ \\
$[$\ion{Cl}{2}$]$   & $14.3678$ & $  1.25$ & $ 0.067$ & $  1.246$ & $-0.535$ & \nodata   &  \nodata & \nodata   & $-0.750$ \\
$[$\ion{Ar}{2}$]$   & $ 6.9853$ & $  14.5$ & $ 0.088$ & $ 9.81$ & $-1.091$ & \nodata   &  \nodata & \nodata   & $-1.309$ \\
$[$\ion{Ar}{3}$]$   & $ 8.9910$ & $ 35.6$ & $ 0.068$ & $ 86.73$ & $-0.089$ & \nodata   &  \nodata & \nodata   & $-0.172$ \\
$[$\ion{Ar}{3}$]$   & $21.8302$ & $  5.82$ & $ 0.121$ & $ 5.19$ & $-0.089$ & $ 4.626$  &  $0.121$ & $  4.62$ & $-0.172$ \\
$[$\ion{Ar}{5}$]$   & $13.1022$ & \nodata   & \nodata  & \nodata   & $-2.796$ & $ 6.031$  &  $0.101$ & $  20.38$ & $-1.276$ \\ 
$[$\ion{Fe}{3}$]$   & $22.9250$ & $  4.35$ & $ 0.145$ & $ 4.34$ & $-0.587$ & $ 17.67$  &  $0.110$ & $  17.63$ & $-0.669$ \\
$[$\ion{Fe}{6}$]$   & $14.7710$ & \nodata   & \nodata  & \nodata   & $-3.199$ & $ 0.446$  &  $0.139$ & $  14.32$ & $-1.207$
\enddata
\tablenotetext{a}{Extinction-corrected observed line fluxes with respect to Br~$\gamma$ (GNIRS) from Table~1 or H~I 7--6 12.37~\micron\ \citep[IRS,][]{simpson:18}. }
\tablenotetext{b}{Uncertainties in the observed line flux ratios, consisting of the statistical uncertainties from the line flux measurements combined quadratically with the uncertainty of the associated hydrogen recombination line.}
\tablenotetext{c}{Modeled line fluxes with respect to Br~$\gamma$ or H~I 7--6 12.37~\micron.}
\tablenotetext{d}{Logarithm of the fractional ionization of each ion.}
\end{deluxetable*}

%% file: tab6.tex

\begin{deluxetable}{lccc}
\tablecaption{Elemental Abundances\label{tab:tab6}}
\tablehead{
   \colhead{Elements} &
   \colhead{SSTGC~580183} & 
   \colhead{SSTGC~588220} &
   \colhead{Solar\tablenotemark{\scriptsize b}}
}
\startdata
Hydrogen  & $12.00        $ & $12.00         $ & $12.00$ \\
Helium & $11.16 \pm 0.04 $ & $10.86 \pm 0.03 $ & $[10.93]$ \\
Oxygen  & $9.52 \pm 1.85 $ & $8.95 \pm 0.35  $ & $8.69$ \\
Neon & $8.64 \pm 0.04 $ & $8.67 \pm 0.04  $ & $[7.93]$ \\
Sodium & $6.84 \pm 0.06 $ & \nodata           & $6.21$ \\
Magnesium & \nodata          & $8.05 \pm 1.29  $ & $7.60$ \\
Phosphorus & $5.78 \pm 0.16 $ & $5.86 \pm 0.16  $ & $5.41$ \\
Sulfur  & $7.59 \pm 0.18 $ & $7.36 \pm 0.22  $ & $7.12$ \\
Chlorine & $5.75 \pm 0.10 $ & \nodata           & $5.50$ \\
Argon & $7.13 \pm 0.04 $ & $6.99 \pm 0.08  $ & $[6.40]$ \\
Iron & $6.99 \pm 0.19 $ & $7.53 \pm 0.15  $ & $7.47$ \\
Krypton\tablenotemark{\scriptsize c} & $4.16 \pm 0.20 $ & $3.67\pm 0.06 $ & $[3.25]$
\enddata
\tablenotetext{a}{Solar abundances from \citet{asplund:09} with the exception of Na, Mg, and Fe, which are from \citet{scott:15a,scott:15b}. The brackets around the noble gas abundances indicate that these solar abundances are `indirect photospheric estimates' \citep{asplund:09}.}
\tablenotetext{b}{Krypton was modeled separately from the Cloudy runs (see text).}
\end{deluxetable}